\documentclass[12pt,a4paper]{article}
\usepackage{jcappub}
\usepackage{bm}
\usepackage{indentfirst}
\usepackage{amsmath}
\usepackage{graphicx}
\usepackage{float}
\usepackage{amssymb}
\usepackage{subfigure}
\usepackage{hyperref}
\usepackage{array}
\usepackage{amsthm}
\usepackage{mathrsfs}
\usepackage{color}
\usepackage{comment}
\usepackage[normalem]{ulem}
\usepackage{url} 
\usepackage{orcidlink}
\hypersetup{
	colorlinks=true,
	linkcolor=red,
	citecolor=blue,
}
\allowdisplaybreaks[2]

\usepackage{xcolor}
\usepackage[normalem]{ulem}

\newcommand{\hws}{\textcolor{cyan}{$\mathcal{HT}$: }\bgroup\markoverwith{\textcolor{cyan}{\rule[.5ex]{2pt}{2.5pt}}}\ULon}

\begin{document}

\title{A Catalog-Wide Study of Gravitational-Wave Residuals in GWTC-4 and GWTC-5}

\author[a]{Junlin Qin,}
\author[a,1]{Zhan-Feng Mai \note{Corresponding author.}}
\author[b,2]{Dicong Liang,\note{Corresponding author.}}
\author[c]{Hai-Tian Wang,}
\author[a]{En-Wei Liang}

\affiliation[a]{Guangxi Key Laboratory for Relativistic Astrophysics, School of Physical Science and Technology, Guangxi University, Nanning 530004, China}
\affiliation[b]{Department of Mathematics and Physics, School of Biomedical Engineering, Southern Medical University, Guangzhou, 510515, China}
\affiliation[c]{School of Physics, Dalian University of Technology, Dalian 116024, China}

\emailAdd{junlin.qin@st.gxu.edu.cn}
\emailAdd{zf1102@gxu.edu.cn }
\emailAdd{dcliang@smu.edu.cn}
\emailAdd{wanght9@dlut.edu.cn}
\emailAdd{lew@gxu.edu.cn }

\abstract{
Improved sensitivity of the LIGO–Virgo–KAGRA detectors has enabled the accumulation of a large gravitational-wave event catalog, providing an opportunity for catalog-wide residual analyses to assess waveform-model performance. 
We study GWTC-4 and GWTC-5 events with signal-to-noise ratio $>12$ in at least one detector, applying three goodness-of-fit tests to the normalized Q-transform energies of whitened residuals obtained by subtracting the IMRPhenomXPHM and SEOBNRv5PHM waveform models. Across the catalog, the residuals are consistent with Gaussian noise, with no evidence of systematic deviations in either multi-detector events or valid single-detector events. For events exhibiting strong model dependence in parameter estimation, variations between waveform models likewise do not produce persistent residual structures. These results demonstrate that parameter-estimation biases associated with waveform modeling do not necessarily manifest as detectable features in the residuals. Residual analyses therefore provide a complementary diagnostic in data space for validating gravitational-wave waveform models within the detector sensitivity.
}

\keywords{gravitational waves; residual test; waveform models; goodness-of-fit tests; GWTC-4; GWTC-5}

\maketitle
%----------------------------------------
\section{Introduction}
%----------------------------------------

Since the first direct detection of gravitational waves (GWs) by LIGO in 2015~\cite{LIGOScientific:2016aoc}, GW astronomy has entered an era of precise observations.
With the continuing improvement in the sensitivity of ground-based GW detectors, the Gravitational-Wave Transient Catalogs (GWTCs) has grown to include hundreds of compact binary coalescence events, providing a wealth of observational data for studying the formation and evolution of compact binary.
With the rapidly expanding sample of detected events, catalog-wide statistical analyses have become an increasingly important aspect of GW data analysis. 
In particular, robust tests of general relativity (GR) should extend beyond a smalll number of representative events to encompass the full catalog of GW detections.
Catalog-wide analyses can combine information across events, reveal systematic trends that may not be apparent in individual observations, and provide more statistically robust constraints on potential deviations from GR.
Notably, the recently released catalogs, GWTC-4~\cite{LIGOScientific:2025hdt,LIGOScientific:2025slb} and GWTC-5~\cite{LIGOScientific:2026wfs}, contain substantially more high signal-to-noise ratio (SNR) events than previous catalogs, enabling more comprehensive investigations of possible deviations from GR.

Current tests of GR using GW observations employ a broad range of complementary approaches \cite{LIGOScientific:2016lio,LIGOScientific:2018dkp,LIGOScientific:2019fpa,PhysRevD.103.122002,LIGOScientific:2021sio,LIGOScientific:2026qni,LIGOScientific:2026fcf,LIGOScientific:2026oim}. 
These include
tests of GW generation and propagation (e.g. \cite{Will:1997bb,Yunes:2009ke,Cornish:2011ys,Mirshekari:2011yq,Wang:2021ctl,Wang:2021gqm,Mehta:2022pcn,LIGOScientific:2026uyd}), 
post-merger and ringdown tests (e.g. \cite{Dreyer:2003bv,Wang:2021uuh,LIGOScientific:2025wao,LIGOScientific:2026wpt}),
polarization tests (e.g. \cite{Eardley:1973br,Eardley:1973zuo,Takeda:2018uai,Takeda:2020tjj,Pang:2020pfz,Wong:2021cmp,Zhang:2021fha,Hu:2023soi,Jiang:2025abg}),
residual tests (e.g. \cite{Johnson-McDaniel:2021yge,Nielsen:2018bhc,Marcoccia:2020rag,Liang:2025zws, Liang:2026nij} ), 
and so on.
Among these, residual tests do not require a specific parameterization of potential deviations from GR. 
Instead, they search for unmodeled features in the data remaining after subtraction of the best-fit waveform, thereby providing a relatively general test of consistency between the observed signal and waveform model. 
For a GW observation, the data recorded by a detector, $d(t)$, can be expressed as the superposition of the GW signal, $s(t)$, and detector noise, $n(t)$:
\begin{equation}
    d(t)=s(t)+n(t).    
\end{equation}
A residual test then subtracts the best-fit waveform, $h(t)$, from the observed data to obtain the residual
\begin{equation}
    r(t)=d(t)-h(t)=n(t)+s(t)-h(t).    
\end{equation}
Real GW detector data are affected by a variety of instrumental and environmental noise sources. 
Over limited time intervals and frequency ranges, the detector noise can often be approximated as a stationary Gaussian process \cite{LIGOScientific:2019hgc,LIGO:2021ppb,LIGO:2024kkz,LIGO:2026ika}.
However, the real data may depart from this idealized noise model because of short-duration transient artifacts, 
narrow-band spectral lines and combs, and persistent
broadband artifacts~\cite{LIGOScientific:2016gtq,Zevin:2016qwy,Glanzer:2022avx,Nuttall:2018xhi,O4LIGODetector:2026okh}. 
If the waveform model accurately describes the true GW signal, the residual after waveform subtraction should be dominated by detector noise, and should exhibit statistical properties consistent with the adopted noise model. 
Conversely, significant structure remaining in the residual may arise from waveform-modeling inaccuracies~\cite{Gupte:2024jfe,Jan:2025zcm,Cole:2022ucw,Zwick:2024yzh,Zwick:2025wkt,Samajdar:2021egv,Relton:2021cax}, insufficiently modeled instrumental noise, or other data-quality issues~\cite{Powell:2018csz,Hourihane:2022doe,Narola:2024qdh}. 
After these possibilities have been carefully assessed, any remaining unexplained residual structure may motivate further investigation into potential deviations from GR.

Recently,  LIGO-Virgo-KAGRA (LVK) Collaboration has used {\tt BayesWave} algorithm~\cite{Cornish:2014kda,Cornish:2020dwh} to performed catalog-wide residual analyses of events GWTC-4 and GWTC-5 \cite{LIGOScientific:2026qni,LIGOScientific:2026oim}. 
{\tt BayesWave} uses a template-independent wavelet model to coherently reconstruct residual features across the detector network.
In this work, we complement these studies by adopting a distinct statistical framework based on classical goodness-of-fit tests applied to the distribution of whitened Q-transformed residual energies.
This framework builds on the approach introduced by \citet{Liang:2025zws}, who applied the Kolmogorov--Smirnov (KS), Anderson--Darling (AD), and chi-squared ($\chi^2$) tests to residuals from GWTC-3. 
Unlike methods based on coherent residual reconstructions, our approach assess the statistical properties of the whitened Q-transformed residual energy separately for each detector. 
It therefore provides an alternative, detector-level characterization of residuals without requiring coherent reconstruction across multiple detectors. 

Our analysis comprises three main components. 
First, we conduct a unified catalog-wide residual analysis of events in GWTC-4 and GWTC-5 that satisfy our SNR selection criteria.
We perform the residual tests using two waveform families, IMRPhenomXPHM~\cite{Pratten:2020ceb} and SEOBNRv5PHM~\cite{Ramos-Buades:2023ehm}, allowing us to assess the impact of waveform-model systematics.
Second, we analyze the single-detector events, which cannot be studied within the coherent {\tt BayesWave} framework used in the previous analyses.
Finally, for events whose parameter-estimation results exhibit pronounced waveform-model dependence, we further examine the residuals obtained from multiple high-likelihood waveforms, to evaluate the robustness of our conclusions.

This paper is organized as follows. 
Sec.~\ref{sec2} describes the data processing, waveform reconstruction, and statistical methods employed in the residual analysis.
In Sec.~\ref{sec3.1}, we use a representative event to illustrate the statistical properties of the signal and residual.
Sec.~\ref{sec3.2} presents the catalog-wide results of residual tests for GWTC-4 and GWTC-5, while Sec.~\ref{sec3.3} focuses on single-detector events.
We then apply ensemble-based residual tests to the events exhibiting significant waveform-model dependence in their parameter-estimation results in Sec.~\ref{sec3.4}. 
Finally, Sec.~\ref{sec4} summarizes our main findings and discusses their implications.

%--------------------------------------------------------
\section{Data and Residual Analysis Framework}
\label{sec2}
%--------------------------------------------------------
The aim of this work is to perform a uniform residual analysis of the latest GW event catalogs.
This requires a sufficiently large event sample together with a consistent data-processing and statistical-analysis pipeline. 
During the fourth observing run (O4), the LVK Collaboration released the GWTC-4 and GWTC-5 catalogs, which contain a substantially expanded sample of compact binary coalescence events compared with previous catalogs. 
In particular, the increased number of high-SNR observations makes it possible to carry out a more comprehensive statistical study at the catalog level.

We use the publicly available strain data and parameter-estimation results from the Gravitational Wave Open Science Center (GWOSC)~\cite{Trovato:2019liz} \footnote{\url{https://gwosc.org/}} for events in GWTC-4 and GWTC-5 that satisfy our SNR selection criteria. 
For each event, the gravitational waveform is reconstructed from the corresponding parameter-estimation results and subsequently subtracted from the detector data to construct the residual. 
The resulting residuals are then assessed within a common statistical framework to determine their compatibility with the Gaussian-noise model.

%------------------------------------------------------
\subsection{Data selection and waveform subtraction}
%------------------------------------------------------

We select events from GWTC-4 and GWTC-5 with a matched-filter SNR greater than 12 in at least one detector. 
We adopt this selection criterion throughout the paper because our residual tests have limited sensitivity to distinguish signals from residuals below this threshold~\cite{Liang:2025zws}.
After applying this selection, our final sample comprises
$N_{\rm events}=\textbf{38}$ events: 
$N_{\rm single}=\textbf{10}$ single-detector events and $N_{\rm multi}=\textbf{28}$ multi-detector events.
For each selected event, the residual analysis is performed independently for every detector that satisfies the SNR requirement.
We restrict our analysis to data from the LIGO Hanford (H1) and LIGO Livingston (L1) detectors because the corresponding Virgo data generally fall below our SNR threshold.

For each event, the GW signal is reconstructed with the IMRPhenomXPHM and SEOBNRv5PHM waveform models, both of which incorporate spin-precession and higher-order-mode effects. 
The source parameters of the maximum-likelihood sample from the corresponding parameter-estimation results are used for the waveform reconstruction.

For each detector, we analyze a 32-second segment of strain data centered on the event time, covering 16 seconds before and after the event, with a sampling rate of 4096~Hz. The noise power
spectral density is estimated from the full strain segment using
median averaging of 4-second Hann-windowed segments with 2-second overlap. 
The resulting noise spectrum is used to apply the same whitening filter to both the observed strain and the reconstructed waveform.
The whitened residual is then obtained by subtracting the reconstructed waveform from the observed data
\begin{equation}
    r_w(t) = d_w(t) - h_w(t),
\end{equation}
where $d_w(t)$ and $h_w(t)$ denote the whitened observed strain and reconstructed waveform, respectively. 
If the waveform model adequately captures the observed signal, the residual after waveform subtraction should contain no significant coherent structure associated with the GW signal, and its statistical properties should be compatible with the whitened noise of the corresponding detector. 
This expectation is quantitatively examined below through time-frequency analysis and goodness-of-fit tests.

%-----------------------------------------------------
\subsection{Time-frequency residual analysis}
%-----------------------------------------------------

The residual obtained after waveform subtraction consists primarily of detector noise, together with any potentially unmodeled signal components. Since non-Gaussian features may be localized within particular regions of time or frequency, a direct time-domain examination may not effectively reveal such structures. 
We therefore extend the analysis to the time-frequency domain, where the energy distribution provides a convenient characterization of the statistical behavior of the residuals.

We perform the time-frequency analysis using the Q-transform implemented in GWpy~\cite{MACLEOD2021100657}, with the corresponding time-frequency energies calculated from the q-gram. 
By employing windows with different time and frequency scales, the Q-transform provides an adaptive trade-off between temporal and frequency resolution. 
This property makes it well suited for characterizing localized features in transient GW signals and has led to its widespread use in GW visualization and excess-power searches~\cite{Chatterji:2004qg,Vazsonyi:2022jul}. 
For a continuous time series, the Q-transform is defined as~\cite{Chatterji:2004qg,Chatterji}
\begin{equation}
    X(\tau,f) = \int_{-\infty}^{\infty}x(t)w(t-\tau,f,Q)e^{-i2\pi ft}\mathrm{d}t,
\end{equation}
where $w(t-\tau,f,Q)$ is a time-domain window centered at $\tau$, whose duration is proportional to $Q$ and inversely proportional to the frequency $f$. For discrete GW data, the corresponding discrete Q-transform is given by
\begin{equation}
    X[m,k] = \sum_{n=0}^{N-1}x[n]w[n-m,k]e^{-i2\pi nk/N},
\end{equation}
where $m$ and $k$ are discrete indices corresponding to time $\tau$ and frequency $f$, respectively. 
In practice, we consider a time interval centered on the merger and restrict the frequency range to $30$--$1000~\mathrm{Hz}$. 
The same time window, frequency range, and q-gram parameters are applied to both
the signal and residual data for each event to ensure a consistent statistical comparison. 
To account for variations in the noise level across the time-frequency plane, the energy is normalized as $y=|X|^2/\langle |X|^2\rangle$, where $\langle |X|^2\rangle$ denotes the arithmetic mean of all nonzero energy values within the selected time-frequency region. 
For whitened Gaussian noise, the normalized Q-transform energy is expected to follow an exponential distribution, $f(y)=e^{-y}$~\cite{Vazsonyi:2022jul,Blackburn:2008ah,Chatterji}.
We therefore adopt the exponential
distribution as the null hypothesis and use goodness-of-fit tests to
quantify the compatibility of the residual energy distribution with
the expectation for Gaussian noise.

%-------------------------------------------------
\subsection{Statistical methods}
%-------------------------------------------------
To assess whether the normalized Q-transform energy distribution matches the theoretical expectation for Gaussian noise, we use three goodness-of-fit tests: the KS, AD, and $\chi^2$ tests.
For all three tests, the $p$-value is defined as the probability, under the null hypothesis, of obtaining a test statistic at least as large as the observed value. 
Throughout this work, we adopt a significance level of $\alpha=0.01$.

%--------------------------------------
\subsubsection{Kolmogorov–Smirnov test}
%--------------------------------------
The KS test~\cite{Kolmogorov,N.-Smirnov} is a nonparametric goodness-of-fit test that quantifies the maximum discrepancy between an empirical distribution function and a reference distribution. For the normalized Q-transform energies considered in this work, the cumulative distribution function (CDF) under the ideal Gaussian-noise hypothesis is $F(y)=1-e^{-y}$. 
We denote the empirical cumulative distribution function of the observed normalized energy samples by $F_N(y)$. 
The KS statistic is then defined as the maximum absolute difference between the empirical and theoretical distributions,
\begin{equation}
    D_N=\sup\limits_y\left|F_N(y)-F(y)\right|,
\end{equation}
A larger value of $D_N$ indicates a greater discrepancy between the observed energy distribution and the theoretical noise model.

%--------------------------------------------
\subsubsection{Anderson–Darling test}
%--------------------------------------------

The AD test provides a complementary measure of goodness of fit by introducing a weighting factor related to $F(x)[1-F(x)]$, which makes it more sensitive to discrepancies in the tails of the distribution than the KS test. 
The AD statistic $A^2$ is defined as~\cite{T.-W.-Anderson}
\begin{equation}
    A^2 = -N-\frac{1}{N}\sum_{i=1}^N(2i-1)
    [\ln F(Y_i) + \ln (1-F(Y_{N+1-i}))],
\end{equation}
where $Y_i$ denotes the energy samples arranged in ascending order, and $F(Y_i)=1-e^{-Y_i}$ is the cumulative distribution function of the exponential distribution. 
For the AD test, the $p$-value is estimated through Monte Carlo simulations and is defined as
\begin{equation}
    p = P\left(A^2_{\rm sim} \geq A^2_{\rm data}\mid H_0\right),
\end{equation}
where the probability is estimated from the simulated null distribution.

%--------------------------------------------------
\subsubsection{Chi-squared test}
%------------------------------------------------------

Unlike the KS and AD tests, the $\chi^2$ test~\cite{pearson1900x} assesses the agreement between the sample distribution and the theoretical model by comparing the observed and expected frequencies in a set of intervals. 
For the normalized Q-transform energies considered here, the samples are divided into a number of energy bins, and the observed frequency $O_i$ is determined for each bin. 
Under the exponential-distribution hypothesis, the corresponding expected frequency $E_i$ can be calculated from the theoretical distribution. 
The test statistic is defined as
\begin{equation}
    \chi^2 = \sum_{i=1}^k \dfrac{(O_i - E_i)^2}{E_i},
\end{equation}
where $k$ is the number of bins, and $O_i$ and $E_i$ denote the observed and expected frequencies in the $i$th energy bin, respectively. 
When the residual energy distribution is consistent with the exponential distribution, the observed frequencies are expected to be close to their theoretical values, resulting in a relatively small $\chi^2$ statistic. 
To avoid unreliable statistics caused by excessively small expected frequencies, the binning is chosen such that the expected frequency in each bin satisfies the required statistical criterion. 
In this work, the normalized energies are divided into 50 bins, with the expected frequencies calculated from the theoretical exponential distribution.

%------------------------------------------------------
\section{Residual Test Results}\label{sec3}
%------------------------------------------------------

Having established the residual-analysis framework in the previous section, we now apply it to the events in GWTC-4 and GWTC-5 that satisfy our selection criteria.
We begin with a representative high-SNR event to illustrate the time-frequency structure and statistical properties of both the observed signal and the corresponding residuals. 
We then present the results for the full sample and assess catalog-wide consistency of the residuals with the Gaussian-noise hypothesis. 
Finally, we extend the analysis to single-detector events and to the events whose parameter-estimation results exhibit significant waveform-model dependence.

%---------------------------------------------
\subsection{Representative event}
\label{sec3.1}
%----------------------------------------------

To illustrate the statistical properties of the signal and residual, we consider a high-SNR event 
$\text{GW250114\_082203}$ (GW250114)~\cite{LIGOScientific:2025wao,LIGOScientific:2025rid} as a representative example.
\begin{figure}[htbp]
    \centering
    \includegraphics[width=\textwidth]{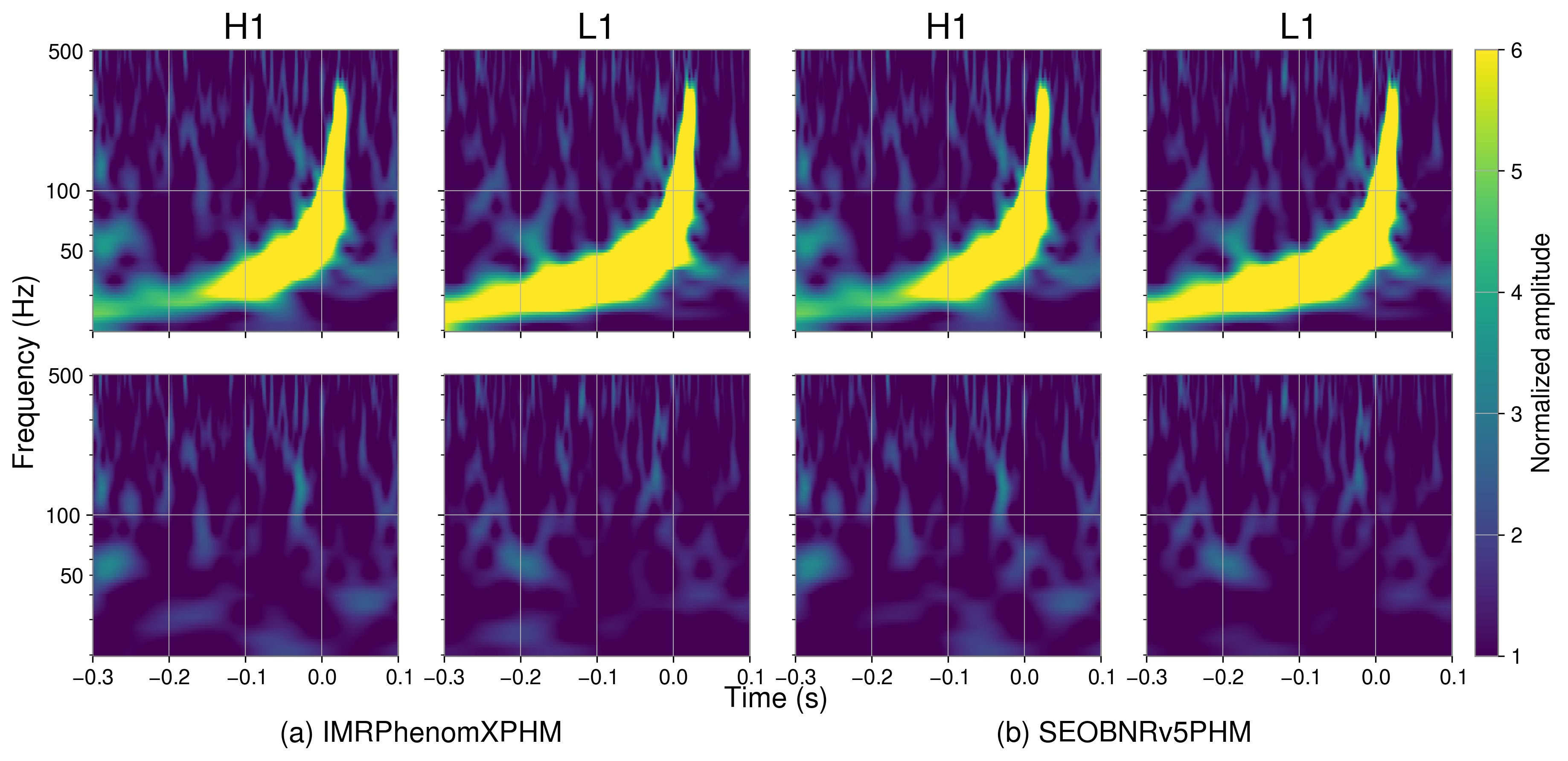}
    \caption{Q-transform spectrograms of the whitened strain (top) and the residuals (bottom) for the H1 and L1 detectors of GW250114 obtained by subtracting the IMRPhenomXPHM and SEOBNRv5PHM waveform models. 
    The horizontal axis shows the relative GPS time, with the merger time at 1420878141.2 s, while the vertical axis represents frequency on a logarithmic scale. }
    \label{qtransform}
\end{figure}
We show the Q-transform time-frequency maps of GW250114 and the corresponding residuals obtained by subtracting the IMRPhenomXPHM and SEOBNRv5PHM waveform models in Fig.~\ref{qtransform}. 
A clear binary-black-hole merger chirp is visible in the original data. 
After the subtraction of the maximum-likelihood waveform, no continuous time-frequency track associated with the original signal appears in the residuals from either model, nor is there any noticeable visual difference between them. 
We next quantify the consistency of the residuals with the noise model by examining the normalized energy distributions and the corresponding statistical tests.
\begin{figure}[htbp]
    \centering
    \includegraphics[width=0.95\textwidth]{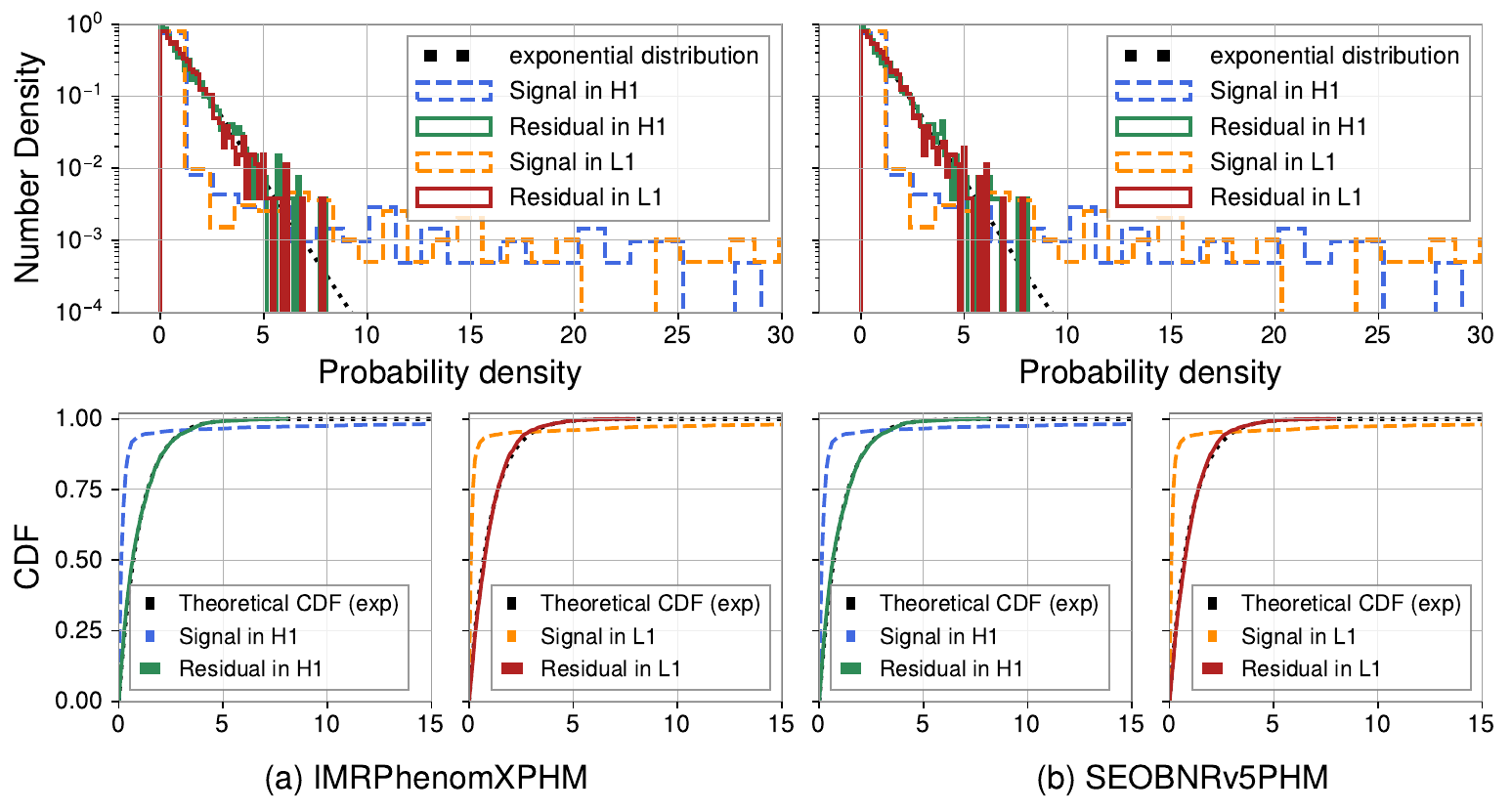}
    \caption{Distributions of the normalized Q-transform energies for signals of $\mathrm{GW}250114$ in the H1 and L1 detectors and the corresponding residuals obtained by subtracting (a) IMRPhenomXPHM and (b) SEOBNRv5PHM. 
    The upper panels show the number density distributions of the signal and residual data compared with the expected exponential distribution, while the lower panels show the corresponding empirical cumulative distribution functions (CDFs) together with the theoretical exponential CDF.}
    \label{hist}
\end{figure}

To compare the statistical properties of the observed signal and the residual more quantitatively, we calculate their normalized Q-transform energy distributions and compare them
with the theoretical exponential distribution, as shown in Fig.~\ref{hist}.
The normalized energy distribution of the original strain departs remarkably from the theoretical exponential distribution at high energies, exhibiting a pronounced long tail that reflects the concentration of gravitational-wave signal power within localized regions of the time-frequency plane. 
In addition, the residual energy distributions after waveform subtraction closely follow the theoretical exponential distribution, and their CDFs also show a good agreement with the theoretical curves. 
For GW250114, both IMRPhenomXPHM and SEOBNRv5PHM effectively remove the dominant signal components from the observed data, leaving no evident residual structure inconsistent with the expectation for whitened Gaussian noise.

%------------------------------------------------
\subsection{Catalog-wide analysis}
\label{sec3.2}
%------------------------------------------------

\begin{figure}[htbp]
    \centering    
    \includegraphics[width=0.98\textwidth]{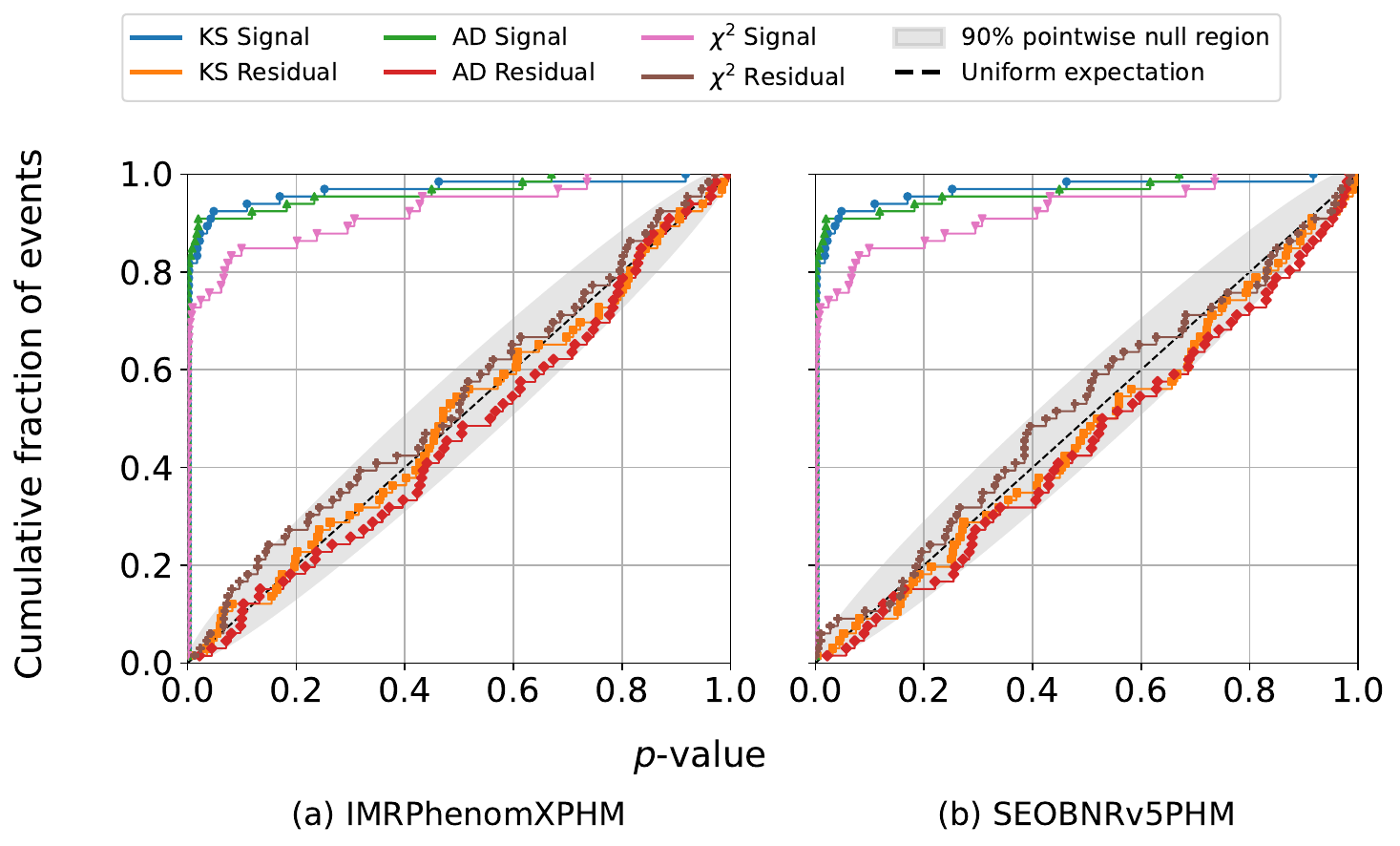}
    \caption{PP-plots of the p-values obtained from the KS, AD, and $\chi^2$ tests for the selected events using (a) IMRPhenomXPHM and (b) SEOBNRv5PHM. The results for the original signals and the residuals are shown separately. The diagonal dashed line represents the theoretical expectation for uniformly distributed p-values, while the shaded region denotes the 90\% pointwise confidence region expected for the order statistics of uniformly distributed $p$-values.}
    \label{pp}
\end{figure}

Now we extend the analysis to all GWTC-4 and GWTC-5 events that satisfy the SNR selection criteria and use the three goodness-of-fit tests to examine the statistical behavior of the residuals under the Gaussian-noise hypothesis.
For each selected event, the waveform is reconstructed using the maximum-likelihood parameters from the IMRPhenomXPHM and SEOBNRv5PHM parameter-estimation posteriors, respectively. 
The corresponding residuals are then calculated. 
The results from all event--detector combinations are then combined in Fig.~\ref{pp}. 
If the residuals are compatible with whitened Gaussian noise, the $p$-values obtained from the goodness-of-fit tests are expected to follow a uniform distribution, $U(0,1)$. 
We summarize the statistical results over the full event sample using probability--probability (PP) plots, which provide a direct comparison between the observed $p$-value distributions and the theoretical expectation for each waveform model.

In a PP plot, the ordered $p$-values are compared with their corresponding theoretical cumulative probabilities. The horizontal axis gives the expected probability positions $i/N$, while the vertical axis shows the ordered values $p_{(i)}$. Under the null hypothesis, the $p$-values follow $U(0,1)$, and the PP curve is therefore expected to lie along $y=x$. To account for finite-sample fluctuations, the $i$th order statistic follows $p_{(i)}\sim\mathrm{Beta}(i, N+1-i)$, from which we construct the 90\% confidence interval.

As shown in Fig.~\ref{pp}, for both the IMRPhenomXPHM and SEOBNRv5PHM waveform models, the test results for the original signals deviate clearly from the theoretical reference line, with a large fraction of the $p$-values concentrated at low values. 
This behavior indicates that the normalized Q-transform energy distributions of the original data differ substantially from the expectation for whitened Gaussian noise. 
After waveform subtraction, in contrast, the PP curves of the residuals lie considerably closer to the expectation for a uniform distribution and remain broadly within the range of finite-sample fluctuations. 
Some differences are still present among the residual PP curves obtained with the three statistical tests, which may reflect both the limited number of events and the different sensitivities of the test statistics to various regions of the underlying distribution. 
To quantify the overall consistency of the residuals with the null hypothesis, we further combine the event-level $p$-values across the catalog using Fisher's combined probability test~\cite{fisher1932statistical}. 
We perform the combination separately for each waveform model and statistical test, yielding six catalog-wide combined $p$-values. The resulting values are summarized in Table~\ref{tab:combined_pvalues}. 
\begin{table}[htbp]
    \centering

    \caption{Catalog-wide combined $p$-values for the three statistical tests applied to the residuals obtained with the IMRPhenomXPHM and SEOBNRv5PHM waveform models.}
    \label{tab:combined_pvalues}

    \vspace{4pt}

    \renewcommand{\arraystretch}{1.25}
    \setlength{\tabcolsep}{14pt}

    \begin{tabular}{ccc}
        \hline
        Statistical test & IMRPhenomXPHM & SEOBNRv5PHM \\
        \hline
        KS       & 0.488 & 0.722 \\
        AD       & 0.849 & 0.928 \\
        $\chi^2$ & 0.155 & 0.100 \\
        \hline
    \end{tabular}
\end{table}
All six combined $p$-values are consistent with the null hypothesis and do not indicate a statistically significant departure at the catalog level.
Overall, within the current event sample and statistical precision, both the PP plots and the catalog-wide combined tests provide no evidence for a significant systematic deviation of the residuals obtained with either waveform model from the whitened Gaussian-noise hypothesis.

%------------------------------------------
\subsection{Single-detector events}
\label{sec3.3}
%------------------------------------------

\begin{figure*}[htbp]
    \centering    
    \includegraphics[width=0.98\textwidth]{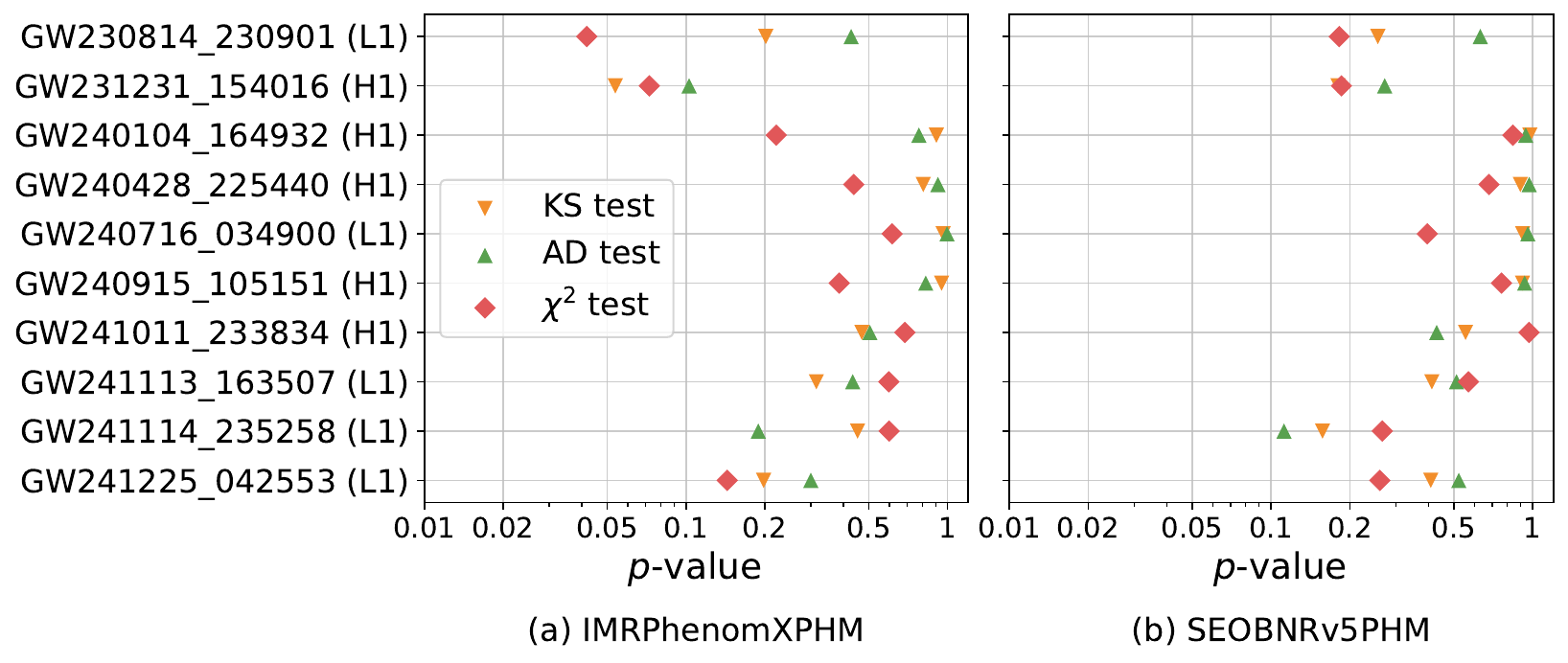}
    \caption{Residual-test results for the selected single-detector events. The left and right panels show the p-values obtained using the (a) IMRPhenomXPHM and (b) SEOBNRv5PHM waveform models, respectively. Different markers denote the KS, AD, and $\chi^2$ tests.}
    \label{event_p}
\end{figure*}

The two catalogs contain several GW events for which only one detector was operating at the time of the event. 
Among them, $N_{\rm H1} = 12$ were observed only by H1, while $N_{\rm L1} = 15$ were observed only by L1 (see Table 1 in Ref.~\cite{LIGOScientific:2025slb} and Table 1 in Ref.~~\cite{LIGOScientific:2026wfs}).
After applying the SNR selection criteria described in Sec.~\ref{sec2}, ten single-detector events remain in our sample.
These events were not included in the previous LVK residual study because coherent reconstruction with {\tt BayesWave} requires data from at least two detectors.
By contrast, our residual analysis can be performed using data from a single detector.  
We present the results for the ten selected single-detector events in in Fig.~\ref{event_p}.

For all the ten events, the resulting $p$-values lie above the adopted significance level, indicating no statistically significant departure of the residuals from the whitened Gaussian-noise hypothesis. 
The three tests produce some variation in the $p$-values for a given event, but their overall statistical conclusions are broadly consistent. Taken together, the results for the single-detector events agree with the catalog-level findings presented above, with no evidence for a systematic departure from the whitened Gaussian-noise hypothesis.

These results also demonstrate that the residual test employed here does not require timing, phase, or amplitude coherence across different detectors. 
Instead, the residual after waveform subtraction can be assessed statistically using data from an individual detector, making the method applicable to GW events for which only one detector record the signal.
It should be emphasized, however, that a single-detector residual test can only determine whether the remaining data are statistically compatible with the adopted noise model; by itself, it cannot identify the origin of a potential anomaly. 
Events showing significant departures therefore require additional investigation incorporating detector data quality, waveform modeling, and other independent tests.

%——-------------------------------------------------------
\subsection{Waveform-model-sensitive events}
\label{sec3.4}
%——-------------------------------------------------------

\begin{figure*}[htbp]
    \centering    
    \includegraphics[width=0.98\textwidth]{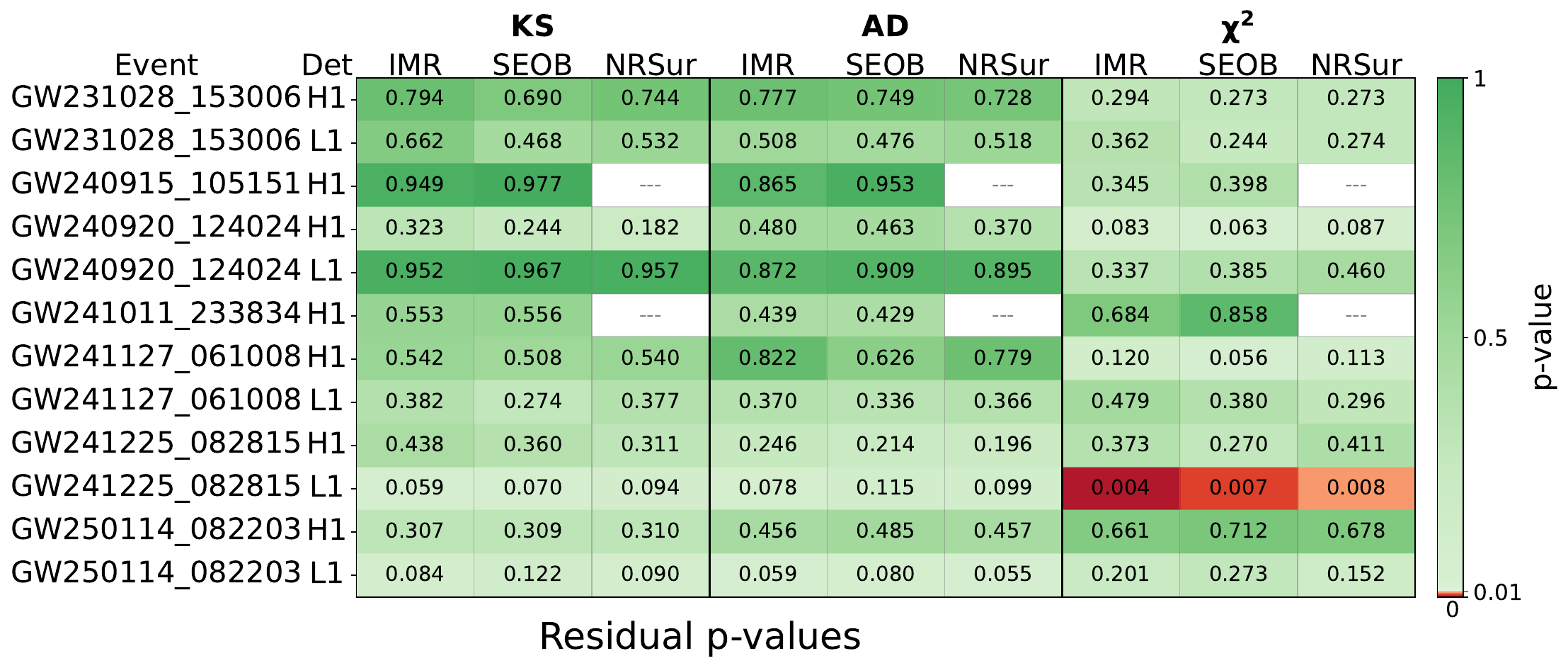}
    \caption{Residual-test $p$-values for the waveform-model-sensitive events, summarized as the median of the $p$-values obtained from the 100 highest-likelihood posterior samples. Rows correspond to event-detector pairs, and columns are grouped by the KS, AD, and $\chi^2$ tests for the IMRPhenomXPHM, SEOBNRv5PHM, and NRSur7dq4 waveform models. The median $p$-values are indicated by both the numerical values and the color scale. A dash denotes an unavailable NRSur7dq4 result.
}
    \label{fig:heatmap}
\end{figure*}

An examination of the catalog-wide parameter-estimation results identifies 14 events whose inferred source parameters exhibit pronounced dependence on the choice of waveform model ~\cite{LIGOScientific:2025slb,LIGOScientific:2026wfs}.
These comprise five events from GWTC-4, summarized in Table 4 of Ref.~\cite{LIGOScientific:2025slb}, and nine events from GWTC-5, summarized in Table 4 of Ref. ~\cite{LIGOScientific:2026wfs}.
Hereafter, we refer to them as waveform-model-sensitive events.
Building on the ensemble-based residual analysis of $\mathrm{GW}231123\_135430$ presented in Ref.~\cite{Liang:2026nij}, we extend this approach to the waveform-model-sensitive events that satisfy our SNR selection criteria.
This analysis allows us to determine whether the waveform-model dependence observed in parameter estimation is also reflected in the residuals derived from ensembles of high-likelihood waveforms.
We select a subset of these waveform-model-sensitive events, according to the SNR selection criteria. 
The selected events are summarized in Fig.~\ref{fig:heatmap}.
Because $\mathrm{GW}231123\_135430$ was already investigated in detail using this approach in our previous study \cite{Liang:2026nij}, we do not revisit it here.

For each selected event, we consider the IMRPhenomXPHM, SEOBNRv5PHM, and NRSur7dq4 \cite{Varma:2019csw} waveform models. 
When posterior samples obtained with NRSur7dq4 are unavailable, we include only the first two models.
For each available waveform model, we select the 100 posterior samples with the highest likelihoods and use them to generate the corresponding waveform realizations and residuals, following the procedure described in Sec.~\ref{sec2}.

The analysis proceeds in two stages. 
We first compare the time-domain residuals associated with different high-likelihood parameter samples within the same waveform model to assess how parameter variations affect waveform subtraction. 
We then apply the KS, AD, and $\chi^2$ tests to the residuals corresponding to the 100 highest-likelihood waveforms for each available waveform model.
The resulting p-values are summarized by their median for each statistical test, providing a representative measure of the consistency between the residuals and the expected noise distribution.

\begin{figure}[htbp]
    \centering

    \begin{subfigure}
        \centering
        \includegraphics[width=\textwidth]{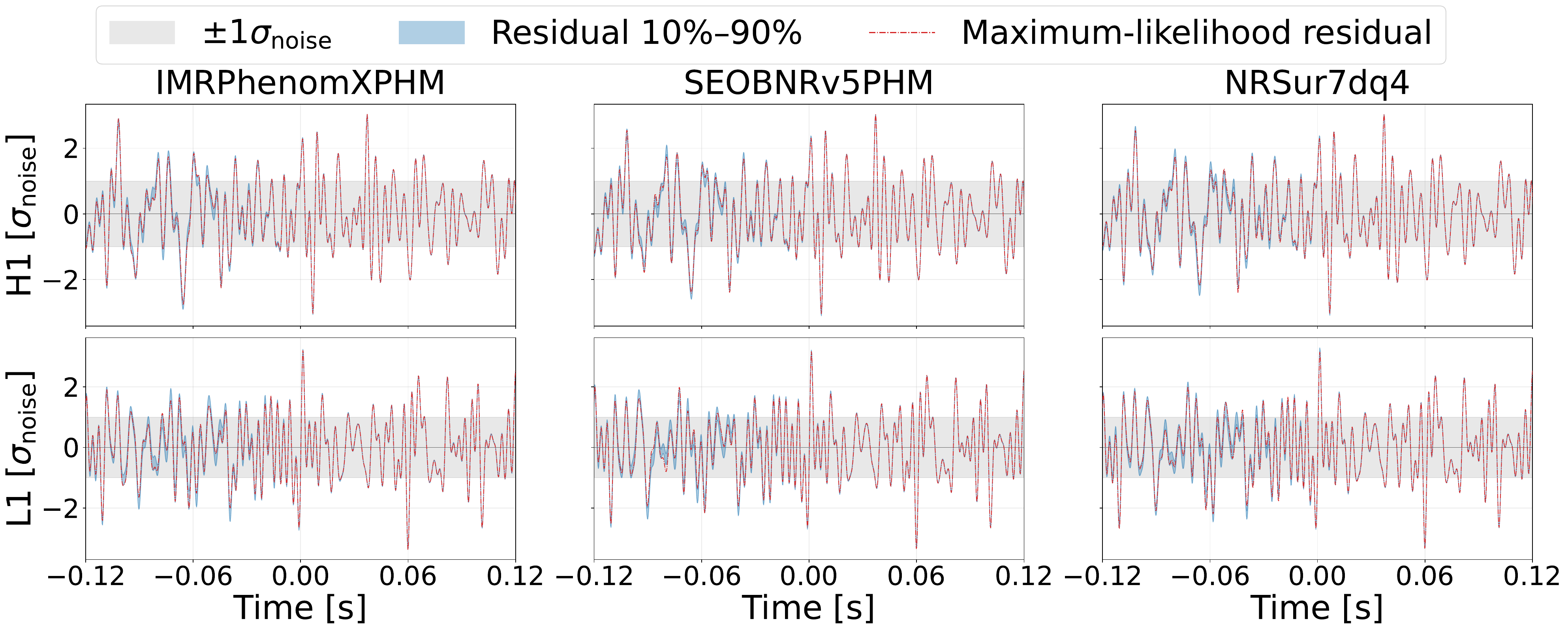}
        \label{fig:GW231028_sigma}
    \end{subfigure}

    \vspace{0.3cm}

    \begin{subfigure}
        \centering
        \includegraphics[width=\textwidth]
        {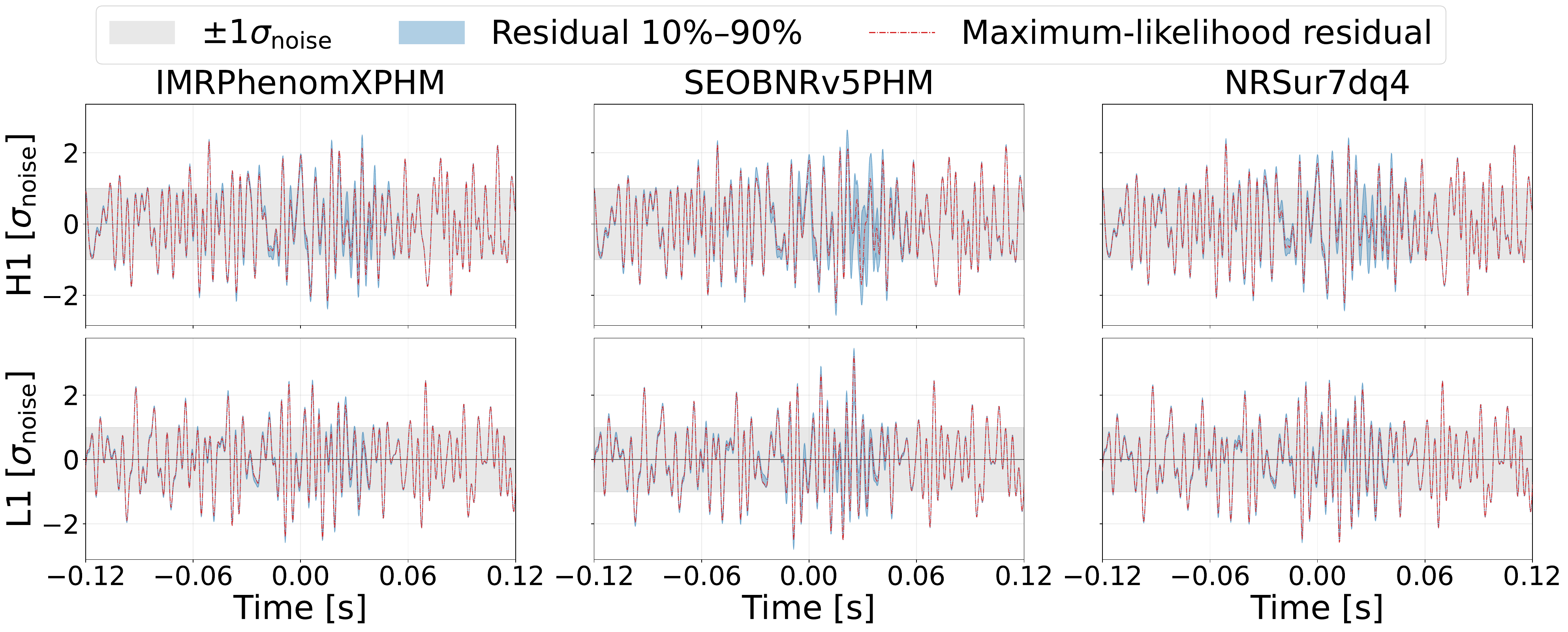}
        \label{fig:GW241127_sigma}
    \end{subfigure}

    \caption{Residuals obtained from the 100 highest-likelihood posterior samples for (a) $\mathrm{GW}231028\_{153006}$ and (b) $\mathrm{GW}241127\_{061008}$, selected as representative waveform-model-sensitive events from O4a and O4b, respectively. The columns correspond to IMRPhenomXPHM, SEOBNRv5PHM, and NRSur7dq4, and the rows to the H1 and L1 detectors. The blue shaded regions show the 10th–90th percentile ranges of the residual realizations, while the red dashed curves denote the maximum-likelihood residuals. The gray shaded regions indicate the ±1$\sigma_{noise}$ levels estimated from the pre-merger off-source data for the corresponding event and detector.
}
    \label{noise}
\end{figure}

Using the high-likelihood waveform realizations described above, we examine in the time domain how variations among the parameter samples affect the resulting residuals, with representative examples shown in Fig.~\ref{noise}. 
To characterize the residual amplitudes relative to the detector background noise, off-source data preceding the merger are used to obtain a robust estimate of the background noise scale through the median absolute deviation (MAD)~\cite{Peter_J}, defined as
\begin{equation}
\sigma_{\mathrm{noise}} = 1.4826 \times \operatorname{median}_{i} \left( \left| x_i - \operatorname{median}_{j}(x_j) \right| \right),
\end{equation}
where $x_i$ denotes the off-source data samples, and the factor 1.4826 rescales the MAD to the standard-deviation scale under the assumption of Gaussian noise. We use $\sigma_{\rm noise}$ as a reference scale for background-noise fluctuations when comparing the residuals obtained from different waveform models and parameter samples.
As shown in Fig.~\ref{noise}, for both events, the residuals associated with high-likelihood parameter samples from the different waveform models generally remain at the level of the background-noise fluctuations, with no persistent or prominent residual structure visible around the merger time. 
The residuals obtained from different high-likelihood samples also overlap substantially, suggesting that the variations induced by parameter uncertainties are relatively limited. Moreover, the residual corresponding to the maximum-likelihood parameters exhibits no anomalous features that distinguish it from those of the other high-likelihood samples, with its time-domain behavior remaining broadly similar to that of the high-likelihood samples considered here.

Although the time-domain comparison provides an intuitive view of the residual structures resulting from different waveform reconstructions, it does not by itself quantify their statistical compatibility with detector noise. 
We therefore apply the three goodness-of-fit tests to the residuals corresponding to the 100 highest-likelihood posterior samples for each waveform model. For each event, waveform model, and statistical test, the resulting 100 $p$-values are summarized by their median, which is used as the representative value for the subsequent comparison.
The corresponding median $p$-values are summarized in Fig.~\ref{fig:heatmap}.

Overall, the vast majority of the resulting $p$-values lie above the adopted significance level. 
Thus, even for events exhibiting pronounced waveform-model dependence in parameter estimation, the residuals obtained from the high-likelihood posterior samples remain broadly compatible with the statistical expectation for whitened Gaussian noise. Although the $p$-values vary to some extent across waveform models, no single model shows a persistent tendency toward systematically lower values across the events considered.

In particular, for $\mathrm{GW}241225\_082815$ in L1, the $\chi^2$ test yields relatively low median $p$-values for all three waveform models. 
Nevertheless, these values remain well above $0.001$, which is approximately quivalent to a one-sided $3\sigma$ significance threshold.
Moreover, the KS and AD tests do not yield comparably low $p$-values for this event. 
Because the three goodness-of-fit tests are sensitive to different aspects of deviations from the expected distribution, a moderately low $p$-value from a single test does not constitute statistically significant evidence against the null hypothesis. 
We therefore interpret the behavior of $\mathrm{GW}241225\_082815$ in L1 as a mild discrepancy detected by the $\chi^2$ test, rather than as robust evidence for anomalous residual structure.

Taken together, the time-domain and statistical results show that differences in waveform reconstruction arising from the choice of waveform model and from variations among high-likelihood parameter samples do not lead to persistent time-domain residual structures or systematic statistical anomalies after waveform subtraction. For the waveform-model-sensitive events examined here, the residual-test results are broadly robust against both waveform-model choice and variations among high-likelihood parameter samples.

%-----------------------------------------------
\section{Summary and Discussion
\label{sec4}}
%-----------------------------------------------

In this work, we perform residual tests on GW events from GWTC-4 and GWTC-5 that satisfy our SNR selection criteria, comparing the IMRPhenomXPHM and SEOBNRv5PHM waveform models. 
The residuals are assessed against the theoretical expectation for Gaussian noise using the KS, AD, and $\chi^2$ tests. 
Within this framework, we systematically examine the statistical compatibility of the residuals with whitened Gaussian noise, while also investigating single-detector events and the effects of waveform-model choice and parameter uncertainties on the residual-test results.

We first use $\text{GW250114}$ as an illustrative example to demonstrate the difference between the statistical properties of signal-containing data and those of the corresponding residuals.
As is shown in Fig.~\ref{hist}, the distribution of normalized Q-transform energies in the original data exhibits a pronounced long tail, indicating a strong departure from the exponential distribution expected for stationary Gaussian noise.
After subtraction of the reconstructed waveform, however, the residual-energy distribution is consistent with the expected exponential distribution, suggesting that the remaining data are compatible with the adopted detector-noise model.
This contrast provides the foundation of our residual tests: signal produces an excess of high-energy time–frequency pixels, whereas an accurate waveform reconstruction should leave residuals whose statistical properties are indistinguishable from those of detector noise. 
Our goodness-of-fit tests quantify any departure from this expected residual distribution.

Next, we calculate the residual $p$-value for each event and detector using three goodness-of-fit tests. 
For each detector and test, we then combine the event-level
$p$-values using Fisher’s combined probability test to assess the overall consistency of the residuals with the null hypothesis.
As is shown in Table~\ref{tab:combined_pvalues}, all six combined $p$-values exceed 0.1.
None of the detector–test combinations provides statistically significant evidence against the null hypothesis.
At the catalog level, the residuals are consistent with the adopted noise model, with no indication of a systematic departure shared across the event population. 
Thus, the current catalog contains no detectable cumulative inconsistency with the null hypothesis.

Then, we focus on ten single-detector events for which data were available from only one detector at the time of observation.
Because coherent reconstruction requires information from at least two detectors, these events cannot be analyzed using conventional coherent multi-detector methods.
Our approach, by contrast, evaluates the residuals independently in each detector and therefore remains fully applicable without relying on cross-detector coherence.
This capability extends residual testing to events that are inaccessible to coherent analyses and constitutes a key advantage of our method over existing network-based residual tests. 
For all ten events, the resulting $p$-values are consistent with the null hypothesis, which are summarized in Fig.~\ref{event_p}. 
We find no statistically significant evidence for unmodeled residual structure. 
Although this does not exclude deviations below the sensitivity of our tests, it demonstrates that single-detector observations can be incorporated into systematic catalog-wide residual analyses.

Finally, we apply ensemble-based residual tests to a set of waveform-model-sensitive events, with the results summarized in Fig.~\ref{fig:heatmap}.
For each waveform model, we use the 100 highest-likelihood posterior samples to reconstruct the waveforms and the corresponding residuals. 
By accounting for variations among the high-likelihood waveforms rather than relying only on a single best-fit waveform, this ensemble-based approach provides a more robust assessment of the residuals. 

Although these events exhibit pronounced waveform-model dependence in parameter estimation, our additional analysis shows that differences between parameter posteriors do not necessarily translate into significant differences at the residual level. 
Different waveform models may favor distinct source parameters while providing comparably adequate descriptions of the signals.
At the current detector sensitivity and statistical precesion, we find no significant evidence that any of the considered model fails to capture the dominant signal content. 
These results demonstrate that waveform-model dependence in parameter space is not fully equivalent to waveform inadequacy in data space. 
Residual tests therefore provide a complementary data-space diagnostic for assessing waveform models beyond direct comparisons of parameter posteriors.

It should be emphasized that the residual tests employed in this work are designed primarily to determine whether the data remaining after waveform subtraction are statistically compatible with the adopted noise model.
It check the consistence between the signal and the waveform model.
Furthermore, residuals alone cannot establish the physical origin of a potential anomaly.
The present analysis is also limited by the number of available events, detector sensitivity, and waveform-model accuracy. 
Subtle waveform-modeling differences may remain hidden by detector noise and may therefore be too weak to produce statistically identifiable features in the residuals. 

As the number of observed events continues to grow and next-generation ground-based GW detectors, such as the Einstein Telescope (ET)~\cite{Punturo:2010zz,ET:2019dnz} and Cosmic Explorer (CE)~\cite{Reitze:2019iox}, come into operation, an increasingly larger population of high-SNR events is expected to become available~\cite{Hild:2010id}. 
The low computational cost of our methods will allow them to be applied efficiently to these rapidly expanding catalogs. 
The increased number and quality of observations will, in turn, enhance the statistical power of residual tests, improving their sensitivity to subtle waveform systematics and weak residual structures.
Combining catalog-level residual statistics with increasingly accurate waveform models may therefore provide a more stringent framework for waveform-model validation and the identification of potential anomalies in future high-precision GW observations.

\section*{Acknowledgments}
%---------------------------------------------------------------------

J. Qin and Z.-F. Mai are supported by the ``Hangji" Action Plan (Guangxi Basic Research Program, Grant No. 2026GXNSFBA00640240), the Guangxi Science and Technology Innovation Platform Program (Leitai Action Plan, Grant No. Guike LT2600640026), the ``Guangxi Highland of Innovation Talents" Program and  the National Natural Science Foundation of China (12605091).
D. Liang is supported by the National Natural Science Foundation of China (12405065, 12465013).
H.-T. Wang is supported by ``the Natural Science Foundation of Liaoning Province" (Grant No. 2025-BS-0065) and ``the Fundamental Research Funds for the Central Universities" at Dalian University of Technology and  the National Natural Science Foundation of China (12603066). 
This research made use of data and software obtained from the Gravitational Wave Open Science Center, a service of LIGO Laboratory, the LIGO Scientific Collaboration, the Virgo Collaboration and KAGRA.

%\bibliographystyle{apsrev4-1}
% \bibliographystyle{unsrt}
%\bibliography{references}

\begin{thebibliography}{79}%
	\makeatletter
	\providecommand \@ifxundefined [1]{%
		\@ifx{#1\undefined}
	}%
	\providecommand \@ifnum [1]{%
		\ifnum #1\expandafter \@firstoftwo
		\else \expandafter \@secondoftwo
		\fi
	}%
	\providecommand \@ifx [1]{%
		\ifx #1\expandafter \@firstoftwo
		\else \expandafter \@secondoftwo
		\fi
	}%
	\providecommand \natexlab [1]{#1}%
	\providecommand \enquote  [1]{``#1''}%
	\providecommand \bibnamefont  [1]{#1}%
	\providecommand \bibfnamefont [1]{#1}%
	\providecommand \citenamefont [1]{#1}%
	\providecommand \href@noop [0]{\@secondoftwo}%
	\providecommand \href [0]{\begingroup \@sanitize@url \@href}%
	\providecommand \@href[1]{\@@startlink{#1}\@@href}%
	\providecommand \@@href[1]{\endgroup#1\@@endlink}%
	\providecommand \@sanitize@url [0]{\catcode `\\12\catcode `\$12\catcode
		`\&12\catcode `\#12\catcode `\^12\catcode `\_12\catcode `\%12\relax}%
	\providecommand \@@startlink[1]{}%
	\providecommand \@@endlink[0]{}%
	\providecommand \url  [0]{\begingroup\@sanitize@url \@url }%
	\providecommand \@url [1]{\endgroup\@href {#1}{\urlprefix }}%
	\providecommand \urlprefix  [0]{URL }%
	\providecommand \Eprint [0]{\href }%
	\providecommand \doibase [0]{http://dx.doi.org/}%
	\providecommand \selectlanguage [0]{\@gobble}%
	\providecommand \bibinfo  [0]{\@secondoftwo}%
	\providecommand \bibfield  [0]{\@secondoftwo}%
	\providecommand \translation [1]{[#1]}%
	\providecommand \BibitemOpen [0]{}%
	\providecommand \bibitemStop [0]{}%
	\providecommand \bibitemNoStop [0]{.\EOS\space}%
	\providecommand \EOS [0]{\spacefactor3000\relax}%
	\providecommand \BibitemShut  [1]{\csname bibitem#1\endcsname}%
	\let\auto@bib@innerbib\@empty
	%</preamble>
	\bibitem [{\citenamefont {Abbott}\ \emph
		{et~al.}(2016{\natexlab{a}})\citenamefont {Abbott} \emph
		{et~al.}}]{LIGOScientific:2016aoc}%
	\BibitemOpen
	\bibfield  {author} {\bibinfo {author} {\bibfnamefont {B.~P.}\ \bibnamefont
			{Abbott}} \emph {et~al.} (\bibinfo {collaboration} {LIGO Scientific,
			Virgo}),\ }\href {\doibase 10.1103/PhysRevLett.116.061102} {\bibfield
		{journal} {\bibinfo  {journal} {Phys. Rev. Lett.}\ }\textbf {\bibinfo
			{volume} {116}},\ \bibinfo {pages} {061102} (\bibinfo {year}
		{2016}{\natexlab{a}})},\ \Eprint {http://arxiv.org/abs/1602.03837}
	{arXiv:1602.03837 [gr-qc]} \BibitemShut {NoStop}%
	\bibitem [{\citenamefont {Abac}\ \emph
		{et~al.}(2025{\natexlab{a}})\citenamefont {Abac} \emph
		{et~al.}}]{LIGOScientific:2025hdt}%
	\BibitemOpen
	\bibfield  {author} {\bibinfo {author} {\bibfnamefont {A.~G.}\ \bibnamefont
			{Abac}} \emph {et~al.} (\bibinfo {collaboration} {LIGO Scientific, KAGRA,
			VIRGO}),\ }\href {\doibase 10.3847/2041-8213/ae0c06} {\bibfield  {journal}
		{\bibinfo  {journal} {Astrophys. J. Lett.}\ }\textbf {\bibinfo {volume}
			{995}},\ \bibinfo {pages} {L18} (\bibinfo {year} {2025}{\natexlab{a}})},\
	\Eprint {http://arxiv.org/abs/2508.18080} {arXiv:2508.18080 [gr-qc]}
	\BibitemShut {NoStop}%
	\bibitem [{\citenamefont {Abac}\ \emph
		{et~al.}(2026{\natexlab{a}})\citenamefont {Abac} \emph
		{et~al.}}]{LIGOScientific:2025slb}%
	\BibitemOpen
	\bibfield  {author} {\bibinfo {author} {\bibfnamefont {A.~G.}\ \bibnamefont
			{Abac}} \emph {et~al.} (\bibinfo {collaboration} {LIGO Scientific, VIRGO,
			KAGRA}),\ }\href {\doibase 10.3847/2041-8213/ae2c74} {\bibfield  {journal}
		{\bibinfo  {journal} {Astrophys. J. Lett.}\ }\textbf {\bibinfo {volume}
			{1004}},\ \bibinfo {pages} {L22} (\bibinfo {year} {2026}{\natexlab{a}})},\
	\Eprint {http://arxiv.org/abs/2508.18082} {arXiv:2508.18082 [gr-qc]}
	\BibitemShut {NoStop}%
	\bibitem [{\citenamefont {Abac}\ \emph
		{et~al.}(2026{\natexlab{b}})\citenamefont {Abac} \emph
		{et~al.}}]{LIGOScientific:2026wfs}%
	\BibitemOpen
	\bibfield  {author} {\bibinfo {author} {\bibfnamefont {A.~G.}\ \bibnamefont
			{Abac}} \emph {et~al.} (\bibinfo {collaboration} {LIGO Scientific, VIRGO,
			KAGRA}),\ }\href@noop {} {\  (\bibinfo {year} {2026}{\natexlab{b}})},\
	\Eprint {http://arxiv.org/abs/2605.27225} {arXiv:2605.27225 [gr-qc]}
	\BibitemShut {NoStop}%
	\bibitem [{\citenamefont {Abbott}\ \emph
		{et~al.}(2016{\natexlab{b}})\citenamefont {Abbott} \emph
		{et~al.}}]{LIGOScientific:2016lio}%
	\BibitemOpen
	\bibfield  {author} {\bibinfo {author} {\bibfnamefont {B.~P.}\ \bibnamefont
			{Abbott}} \emph {et~al.} (\bibinfo {collaboration} {LIGO Scientific,
			Virgo}),\ }\href {\doibase 10.1103/PhysRevLett.116.221101} {\bibfield
		{journal} {\bibinfo  {journal} {Phys. Rev. Lett.}\ }\textbf {\bibinfo
			{volume} {116}},\ \bibinfo {pages} {221101} (\bibinfo {year}
		{2016}{\natexlab{b}})},\ \bibinfo {note} {[Erratum: Phys.Rev.Lett. 121,
		129902 (2018)]},\ \Eprint {http://arxiv.org/abs/1602.03841} {arXiv:1602.03841
		[gr-qc]} \BibitemShut {NoStop}%
	\bibitem [{\citenamefont {Abbott}\ \emph
		{et~al.}(2019{\natexlab{a}})\citenamefont {Abbott} \emph
		{et~al.}}]{LIGOScientific:2018dkp}%
	\BibitemOpen
	\bibfield  {author} {\bibinfo {author} {\bibfnamefont {B.~P.}\ \bibnamefont
			{Abbott}} \emph {et~al.} (\bibinfo {collaboration} {LIGO Scientific,
			Virgo}),\ }\href {\doibase 10.1103/PhysRevLett.123.011102} {\bibfield
		{journal} {\bibinfo  {journal} {Phys. Rev. Lett.}\ }\textbf {\bibinfo
			{volume} {123}},\ \bibinfo {pages} {011102} (\bibinfo {year}
		{2019}{\natexlab{a}})},\ \Eprint {http://arxiv.org/abs/1811.00364}
	{arXiv:1811.00364 [gr-qc]} \BibitemShut {NoStop}%
	\bibitem [{\citenamefont {Abbott}\ \emph
		{et~al.}(2019{\natexlab{b}})\citenamefont {Abbott} \emph
		{et~al.}}]{LIGOScientific:2019fpa}%
	\BibitemOpen
	\bibfield  {author} {\bibinfo {author} {\bibfnamefont {B.~P.}\ \bibnamefont
			{Abbott}} \emph {et~al.} (\bibinfo {collaboration} {LIGO Scientific,
			Virgo}),\ }\href {\doibase 10.1103/PhysRevD.100.104036} {\bibfield  {journal}
		{\bibinfo  {journal} {Phys. Rev. D}\ }\textbf {\bibinfo {volume} {100}},\
		\bibinfo {pages} {104036} (\bibinfo {year} {2019}{\natexlab{b}})},\ \Eprint
	{http://arxiv.org/abs/1903.04467} {arXiv:1903.04467 [gr-qc]} \BibitemShut
	{NoStop}%
	\bibitem [{\citenamefont {Abbott}\ \emph {et~al.}(2021)\citenamefont {Abbott}
		\emph {et~al.}}]{PhysRevD.103.122002}%
	\BibitemOpen
	\bibfield  {author} {\bibinfo {author} {\bibfnamefont {R.}~\bibnamefont
			{Abbott}} \emph {et~al.} (\bibinfo {collaboration} {LIGO Scientific
			Collaboration and Virgo Collaboration}),\ }\href {\doibase
		10.1103/PhysRevD.103.122002} {\bibfield  {journal} {\bibinfo  {journal}
			{Phys. Rev. D}\ }\textbf {\bibinfo {volume} {103}},\ \bibinfo {pages}
		{122002} (\bibinfo {year} {2021})}\BibitemShut {NoStop}%
	\bibitem [{\citenamefont {Abbott}\ \emph {et~al.}(2025)\citenamefont {Abbott}
		\emph {et~al.}}]{LIGOScientific:2021sio}%
	\BibitemOpen
	\bibfield  {author} {\bibinfo {author} {\bibfnamefont {R.}~\bibnamefont
			{Abbott}} \emph {et~al.} (\bibinfo {collaboration} {LIGO Scientific, VIRGO,
			KAGRA}),\ }\href {\doibase 10.1103/PhysRevD.112.084080} {\bibfield  {journal}
		{\bibinfo  {journal} {Phys. Rev. D}\ }\textbf {\bibinfo {volume} {112}},\
		\bibinfo {pages} {084080} (\bibinfo {year} {2025})},\ \Eprint
	{http://arxiv.org/abs/2112.06861} {arXiv:2112.06861 [gr-qc]} \BibitemShut
	{NoStop}%
	\bibitem [{\citenamefont {Abac}\ \emph
		{et~al.}(2026{\natexlab{c}})\citenamefont {Abac} \emph
		{et~al.}}]{LIGOScientific:2026qni}%
	\BibitemOpen
	\bibfield  {author} {\bibinfo {author} {\bibfnamefont {A.~G.}\ \bibnamefont
			{Abac}} \emph {et~al.} (\bibinfo {collaboration} {LIGO Scientific, VIRGO,
			KAGRA}),\ }\href@noop {} {\  (\bibinfo {year} {2026}{\natexlab{c}})},\
	\Eprint {http://arxiv.org/abs/2603.19019} {arXiv:2603.19019 [gr-qc]}
	\BibitemShut {NoStop}%
	\bibitem [{\citenamefont {Abac}\ \emph
		{et~al.}(2026{\natexlab{d}})\citenamefont {Abac} \emph
		{et~al.}}]{LIGOScientific:2026fcf}%
	\BibitemOpen
	\bibfield  {author} {\bibinfo {author} {\bibfnamefont {A.~G.}\ \bibnamefont
			{Abac}} \emph {et~al.} (\bibinfo {collaboration} {LIGO Scientific, VIRGO,
			KAGRA}),\ }\href@noop {} {\  (\bibinfo {year} {2026}{\natexlab{d}})},\
	\Eprint {http://arxiv.org/abs/2603.19020} {arXiv:2603.19020 [gr-qc]}
	\BibitemShut {NoStop}%
	\bibitem [{\citenamefont {Abac}\ \emph
		{et~al.}(2026{\natexlab{e}})\citenamefont {Abac} \emph
		{et~al.}}]{LIGOScientific:2026oim}%
	\BibitemOpen
	\bibfield  {author} {\bibinfo {author} {\bibfnamefont {A.~G.}\ \bibnamefont
			{Abac}} \emph {et~al.} (\bibinfo {collaboration} {LIGO Scientific, VIRGO,
			KAGRA}),\ }\href@noop {} {\  (\bibinfo {year} {2026}{\natexlab{e}})},\
	\Eprint {http://arxiv.org/abs/2607.19293} {arXiv:2607.19293 [gr-qc]}
	\BibitemShut {NoStop}%
	\bibitem [{\citenamefont {Will}(1998)}]{Will:1997bb}%
	\BibitemOpen
	\bibfield  {author} {\bibinfo {author} {\bibfnamefont {C.~M.}\ \bibnamefont
			{Will}},\ }\href {\doibase 10.1103/PhysRevD.57.2061} {\bibfield  {journal}
		{\bibinfo  {journal} {Phys. Rev. D}\ }\textbf {\bibinfo {volume} {57}},\
		\bibinfo {pages} {2061} (\bibinfo {year} {1998})},\ \Eprint
	{http://arxiv.org/abs/gr-qc/9709011} {arXiv:gr-qc/9709011} \BibitemShut
	{NoStop}%
	\bibitem [{\citenamefont {Yunes}\ and\ \citenamefont
		{Pretorius}(2009)}]{Yunes:2009ke}%
	\BibitemOpen
	\bibfield  {author} {\bibinfo {author} {\bibfnamefont {N.}~\bibnamefont
			{Yunes}}\ and\ \bibinfo {author} {\bibfnamefont {F.}~\bibnamefont
			{Pretorius}},\ }\href {\doibase 10.1103/PhysRevD.80.122003} {\bibfield
		{journal} {\bibinfo  {journal} {Phys. Rev. D}\ }\textbf {\bibinfo {volume}
			{80}},\ \bibinfo {pages} {122003} (\bibinfo {year} {2009})},\ \Eprint
	{http://arxiv.org/abs/0909.3328} {arXiv:0909.3328 [gr-qc]} \BibitemShut
	{NoStop}%
	\bibitem [{\citenamefont {Cornish}\ \emph {et~al.}(2011)\citenamefont
		{Cornish}, \citenamefont {Sampson}, \citenamefont {Yunes},\ and\
		\citenamefont {Pretorius}}]{Cornish:2011ys}%
	\BibitemOpen
	\bibfield  {author} {\bibinfo {author} {\bibfnamefont {N.}~\bibnamefont
			{Cornish}}, \bibinfo {author} {\bibfnamefont {L.}~\bibnamefont {Sampson}},
		\bibinfo {author} {\bibfnamefont {N.}~\bibnamefont {Yunes}}, \ and\ \bibinfo
		{author} {\bibfnamefont {F.}~\bibnamefont {Pretorius}},\ }\href {\doibase
		10.1103/PhysRevD.84.062003} {\bibfield  {journal} {\bibinfo  {journal} {Phys.
				Rev. D}\ }\textbf {\bibinfo {volume} {84}},\ \bibinfo {pages} {062003}
		(\bibinfo {year} {2011})},\ \Eprint {http://arxiv.org/abs/1105.2088}
	{arXiv:1105.2088 [gr-qc]} \BibitemShut {NoStop}%
	\bibitem [{\citenamefont {Mirshekari}\ \emph {et~al.}(2012)\citenamefont
		{Mirshekari}, \citenamefont {Yunes},\ and\ \citenamefont
		{Will}}]{Mirshekari:2011yq}%
	\BibitemOpen
	\bibfield  {author} {\bibinfo {author} {\bibfnamefont {S.}~\bibnamefont
			{Mirshekari}}, \bibinfo {author} {\bibfnamefont {N.}~\bibnamefont {Yunes}}, \
		and\ \bibinfo {author} {\bibfnamefont {C.~M.}\ \bibnamefont {Will}},\ }\href
	{\doibase 10.1103/PhysRevD.85.024041} {\bibfield  {journal} {\bibinfo
			{journal} {Phys. Rev. D}\ }\textbf {\bibinfo {volume} {85}},\ \bibinfo
		{pages} {024041} (\bibinfo {year} {2012})},\ \Eprint
	{http://arxiv.org/abs/1110.2720} {arXiv:1110.2720 [gr-qc]} \BibitemShut
	{NoStop}%
	\bibitem [{\citenamefont {Wang}\ \emph
		{et~al.}(2021{\natexlab{a}})\citenamefont {Wang}, \citenamefont {Shao},\ and\
		\citenamefont {Liu}}]{Wang:2021ctl}%
	\BibitemOpen
	\bibfield  {author} {\bibinfo {author} {\bibfnamefont {Z.}~\bibnamefont
			{Wang}}, \bibinfo {author} {\bibfnamefont {L.}~\bibnamefont {Shao}}, \ and\
		\bibinfo {author} {\bibfnamefont {C.}~\bibnamefont {Liu}},\ }\href {\doibase
		10.3847/1538-4357/ac223c} {\bibfield  {journal} {\bibinfo  {journal}
			{Astrophys. J.}\ }\textbf {\bibinfo {volume} {921}},\ \bibinfo {pages} {158}
		(\bibinfo {year} {2021}{\natexlab{a}})},\ \Eprint
	{http://arxiv.org/abs/2108.02974} {arXiv:2108.02974 [gr-qc]} \BibitemShut
	{NoStop}%
	\bibitem [{\citenamefont {Wang}\ \emph {et~al.}(2022)\citenamefont {Wang},
		\citenamefont {Brown}, \citenamefont {Shao},\ and\ \citenamefont
		{Zhao}}]{Wang:2021gqm}%
	\BibitemOpen
	\bibfield  {author} {\bibinfo {author} {\bibfnamefont {Y.-F.}\ \bibnamefont
			{Wang}}, \bibinfo {author} {\bibfnamefont {S.~M.}\ \bibnamefont {Brown}},
		\bibinfo {author} {\bibfnamefont {L.}~\bibnamefont {Shao}}, \ and\ \bibinfo
		{author} {\bibfnamefont {W.}~\bibnamefont {Zhao}},\ }\href {\doibase
		10.1103/PhysRevD.106.084005} {\bibfield  {journal} {\bibinfo  {journal}
			{Phys. Rev. D}\ }\textbf {\bibinfo {volume} {106}},\ \bibinfo {pages}
		{084005} (\bibinfo {year} {2022})},\ \Eprint
	{http://arxiv.org/abs/2109.09718} {arXiv:2109.09718 [astro-ph.HE]}
	\BibitemShut {NoStop}%
	\bibitem [{\citenamefont {Mehta}\ \emph {et~al.}(2023)\citenamefont {Mehta},
		\citenamefont {Buonanno}, \citenamefont {Cotesta}, \citenamefont {Ghosh},
		\citenamefont {Sennett},\ and\ \citenamefont {Steinhoff}}]{Mehta:2022pcn}%
	\BibitemOpen
	\bibfield  {author} {\bibinfo {author} {\bibfnamefont {A.~K.}\ \bibnamefont
			{Mehta}}, \bibinfo {author} {\bibfnamefont {A.}~\bibnamefont {Buonanno}},
		\bibinfo {author} {\bibfnamefont {R.}~\bibnamefont {Cotesta}}, \bibinfo
		{author} {\bibfnamefont {A.}~\bibnamefont {Ghosh}}, \bibinfo {author}
		{\bibfnamefont {N.}~\bibnamefont {Sennett}}, \ and\ \bibinfo {author}
		{\bibfnamefont {J.}~\bibnamefont {Steinhoff}},\ }\href {\doibase
		10.1103/PhysRevD.107.044020} {\bibfield  {journal} {\bibinfo  {journal}
			{Phys. Rev. D}\ }\textbf {\bibinfo {volume} {107}},\ \bibinfo {pages}
		{044020} (\bibinfo {year} {2023})},\ \Eprint
	{http://arxiv.org/abs/2203.13937} {arXiv:2203.13937 [gr-qc]} \BibitemShut
	{NoStop}%
	\bibitem [{\citenamefont {Abac}\ \emph
		{et~al.}(2026{\natexlab{f}})\citenamefont {Abac} \emph
		{et~al.}}]{LIGOScientific:2026uyd}%
	\BibitemOpen
	\bibfield  {author} {\bibinfo {author} {\bibfnamefont {A.~G.}\ \bibnamefont
			{Abac}} \emph {et~al.} (\bibinfo {collaboration} {LIGO Scientific, VIRGO,
			KAGRA}),\ }\href@noop {} {\  (\bibinfo {year} {2026}{\natexlab{f}})},\
	\Eprint {http://arxiv.org/abs/2605.27227} {arXiv:2605.27227 [astro-ph.CO]}
	\BibitemShut {NoStop}%
	\bibitem [{\citenamefont {Dreyer}\ \emph {et~al.}(2004)\citenamefont {Dreyer},
		\citenamefont {Kelly}, \citenamefont {Krishnan}, \citenamefont {Finn},
		\citenamefont {Garrison},\ and\ \citenamefont
		{Lopez-Aleman}}]{Dreyer:2003bv}%
	\BibitemOpen
	\bibfield  {author} {\bibinfo {author} {\bibfnamefont {O.}~\bibnamefont
			{Dreyer}}, \bibinfo {author} {\bibfnamefont {B.~J.}\ \bibnamefont {Kelly}},
		\bibinfo {author} {\bibfnamefont {B.}~\bibnamefont {Krishnan}}, \bibinfo
		{author} {\bibfnamefont {L.~S.}\ \bibnamefont {Finn}}, \bibinfo {author}
		{\bibfnamefont {D.}~\bibnamefont {Garrison}}, \ and\ \bibinfo {author}
		{\bibfnamefont {R.}~\bibnamefont {Lopez-Aleman}},\ }\href {\doibase
		10.1088/0264-9381/21/4/003} {\bibfield  {journal} {\bibinfo  {journal}
			{Class. Quant. Grav.}\ }\textbf {\bibinfo {volume} {21}},\ \bibinfo {pages}
		{787} (\bibinfo {year} {2004})},\ \Eprint
	{http://arxiv.org/abs/gr-qc/0309007} {arXiv:gr-qc/0309007} \BibitemShut
	{NoStop}%
	\bibitem [{\citenamefont {Wang}\ \emph
		{et~al.}(2021{\natexlab{b}})\citenamefont {Wang}, \citenamefont {Tang},
		\citenamefont {Li},\ and\ \citenamefont {Fan}}]{Wang:2021uuh}%
	\BibitemOpen
	\bibfield  {author} {\bibinfo {author} {\bibfnamefont {H.-T.}\ \bibnamefont
			{Wang}}, \bibinfo {author} {\bibfnamefont {S.-P.}\ \bibnamefont {Tang}},
		\bibinfo {author} {\bibfnamefont {P.-C.}\ \bibnamefont {Li}}, \ and\ \bibinfo
		{author} {\bibfnamefont {Y.-Z.}\ \bibnamefont {Fan}},\ }\href {\doibase
		10.1103/PhysRevD.104.104063} {\bibfield  {journal} {\bibinfo  {journal}
			{Phys. Rev. D}\ }\textbf {\bibinfo {volume} {104}},\ \bibinfo {pages}
		{104063} (\bibinfo {year} {2021}{\natexlab{b}})},\ \Eprint
	{http://arxiv.org/abs/2104.07594} {arXiv:2104.07594 [gr-qc]} \BibitemShut
	{NoStop}%
	\bibitem [{\citenamefont {Abac}\ \emph
		{et~al.}(2026{\natexlab{g}})\citenamefont {Abac} \emph
		{et~al.}}]{LIGOScientific:2025wao}%
	\BibitemOpen
	\bibfield  {author} {\bibinfo {author} {\bibfnamefont {A.~G.}\ \bibnamefont
			{Abac}} \emph {et~al.} (\bibinfo {collaboration} {LIGO Scientific, Virgo,
			KAGRA}),\ }\href {\doibase 10.1103/6c61-fm1n} {\bibfield  {journal} {\bibinfo
			{journal} {Phys. Rev. Lett.}\ }\textbf {\bibinfo {volume} {136}},\ \bibinfo
		{pages} {041403} (\bibinfo {year} {2026}{\natexlab{g}})},\ \Eprint
	{http://arxiv.org/abs/2509.08099} {arXiv:2509.08099 [gr-qc]} \BibitemShut
	{NoStop}%
	\bibitem [{\citenamefont {Abac}\ \emph
		{et~al.}(2026{\natexlab{h}})\citenamefont {Abac} \emph
		{et~al.}}]{LIGOScientific:2026wpt}%
	\BibitemOpen
	\bibfield  {author} {\bibinfo {author} {\bibfnamefont {A.~G.}\ \bibnamefont
			{Abac}} \emph {et~al.} (\bibinfo {collaboration} {LIGO Scientific, VIRGO,
			KAGRA}),\ }\href@noop {} {\  (\bibinfo {year} {2026}{\natexlab{h}})},\
	\Eprint {http://arxiv.org/abs/2603.19021} {arXiv:2603.19021 [gr-qc]}
	\BibitemShut {NoStop}%
	\bibitem [{\citenamefont {Eardley}\ \emph
		{et~al.}(1973{\natexlab{a}})\citenamefont {Eardley}, \citenamefont {Lee},
		\citenamefont {Lightman}, \citenamefont {Wagoner},\ and\ \citenamefont
		{Will}}]{Eardley:1973br}%
	\BibitemOpen
	\bibfield  {author} {\bibinfo {author} {\bibfnamefont {D.~M.}\ \bibnamefont
			{Eardley}}, \bibinfo {author} {\bibfnamefont {D.~L.}\ \bibnamefont {Lee}},
		\bibinfo {author} {\bibfnamefont {A.~P.}\ \bibnamefont {Lightman}}, \bibinfo
		{author} {\bibfnamefont {R.~V.}\ \bibnamefont {Wagoner}}, \ and\ \bibinfo
		{author} {\bibfnamefont {C.~M.}\ \bibnamefont {Will}},\ }\href {\doibase
		10.1103/PhysRevLett.30.884} {\bibfield  {journal} {\bibinfo  {journal} {Phys.
				Rev. Lett.}\ }\textbf {\bibinfo {volume} {30}},\ \bibinfo {pages} {884}
		(\bibinfo {year} {1973}{\natexlab{a}})}\BibitemShut {NoStop}%
	\bibitem [{\citenamefont {Eardley}\ \emph
		{et~al.}(1973{\natexlab{b}})\citenamefont {Eardley}, \citenamefont {Lee},\
		and\ \citenamefont {Lightman}}]{Eardley:1973zuo}%
	\BibitemOpen
	\bibfield  {author} {\bibinfo {author} {\bibfnamefont {D.~M.}\ \bibnamefont
			{Eardley}}, \bibinfo {author} {\bibfnamefont {D.~L.}\ \bibnamefont {Lee}}, \
		and\ \bibinfo {author} {\bibfnamefont {A.~P.}\ \bibnamefont {Lightman}},\
	}\href {\doibase 10.1103/PhysRevD.8.3308} {\bibfield  {journal} {\bibinfo
			{journal} {Phys. Rev. D}\ }\textbf {\bibinfo {volume} {8}},\ \bibinfo {pages}
		{3308} (\bibinfo {year} {1973}{\natexlab{b}})}\BibitemShut {NoStop}%
	\bibitem [{\citenamefont {Takeda}\ \emph {et~al.}(2018)\citenamefont {Takeda},
		\citenamefont {Nishizawa}, \citenamefont {Michimura}, \citenamefont {Nagano},
		\citenamefont {Komori}, \citenamefont {Ando},\ and\ \citenamefont
		{Hayama}}]{Takeda:2018uai}%
	\BibitemOpen
	\bibfield  {author} {\bibinfo {author} {\bibfnamefont {H.}~\bibnamefont
			{Takeda}}, \bibinfo {author} {\bibfnamefont {A.}~\bibnamefont {Nishizawa}},
		\bibinfo {author} {\bibfnamefont {Y.}~\bibnamefont {Michimura}}, \bibinfo
		{author} {\bibfnamefont {K.}~\bibnamefont {Nagano}}, \bibinfo {author}
		{\bibfnamefont {K.}~\bibnamefont {Komori}}, \bibinfo {author} {\bibfnamefont
			{M.}~\bibnamefont {Ando}}, \ and\ \bibinfo {author} {\bibfnamefont
			{K.}~\bibnamefont {Hayama}},\ }\href {\doibase 10.1103/PhysRevD.98.022008}
	{\bibfield  {journal} {\bibinfo  {journal} {Phys. Rev. D}\ }\textbf {\bibinfo
			{volume} {98}},\ \bibinfo {pages} {022008} (\bibinfo {year} {2018})},\
	\Eprint {http://arxiv.org/abs/1806.02182} {arXiv:1806.02182 [gr-qc]}
	\BibitemShut {NoStop}%
	\bibitem [{\citenamefont {Takeda}\ \emph {et~al.}(2021)\citenamefont {Takeda},
		\citenamefont {Morisaki},\ and\ \citenamefont {Nishizawa}}]{Takeda:2020tjj}%
	\BibitemOpen
	\bibfield  {author} {\bibinfo {author} {\bibfnamefont {H.}~\bibnamefont
			{Takeda}}, \bibinfo {author} {\bibfnamefont {S.}~\bibnamefont {Morisaki}}, \
		and\ \bibinfo {author} {\bibfnamefont {A.}~\bibnamefont {Nishizawa}},\ }\href
	{\doibase 10.1103/PhysRevD.103.064037} {\bibfield  {journal} {\bibinfo
			{journal} {Phys. Rev. D}\ }\textbf {\bibinfo {volume} {103}},\ \bibinfo
		{pages} {064037} (\bibinfo {year} {2021})},\ \Eprint
	{http://arxiv.org/abs/2010.14538} {arXiv:2010.14538 [gr-qc]} \BibitemShut
	{NoStop}%
	\bibitem [{\citenamefont {Pang}\ \emph {et~al.}(2020)\citenamefont {Pang},
		\citenamefont {Lo}, \citenamefont {Wong}, \citenamefont {Li},\ and\
		\citenamefont {Van Den~Broeck}}]{Pang:2020pfz}%
	\BibitemOpen
	\bibfield  {author} {\bibinfo {author} {\bibfnamefont {P.~T.~H.}\
			\bibnamefont {Pang}}, \bibinfo {author} {\bibfnamefont {R.~K.~L.}\
			\bibnamefont {Lo}}, \bibinfo {author} {\bibfnamefont {I.~C.~F.}\ \bibnamefont
			{Wong}}, \bibinfo {author} {\bibfnamefont {T.~G.~F.}\ \bibnamefont {Li}}, \
		and\ \bibinfo {author} {\bibfnamefont {C.}~\bibnamefont {Van Den~Broeck}},\
	}\href {\doibase 10.1103/PhysRevD.101.104055} {\bibfield  {journal} {\bibinfo
			{journal} {Phys. Rev. D}\ }\textbf {\bibinfo {volume} {101}},\ \bibinfo
		{pages} {104055} (\bibinfo {year} {2020})},\ \Eprint
	{http://arxiv.org/abs/2003.07375} {arXiv:2003.07375 [gr-qc]} \BibitemShut
	{NoStop}%
	\bibitem [{\citenamefont {Wong}\ \emph {et~al.}(2021)\citenamefont {Wong},
		\citenamefont {Pang}, \citenamefont {Lo}, \citenamefont {Li},\ and\
		\citenamefont {Van Den~Broeck}}]{Wong:2021cmp}%
	\BibitemOpen
	\bibfield  {author} {\bibinfo {author} {\bibfnamefont {I.~C.~F.}\
			\bibnamefont {Wong}}, \bibinfo {author} {\bibfnamefont {P.~T.~H.}\
			\bibnamefont {Pang}}, \bibinfo {author} {\bibfnamefont {R.~K.~L.}\
			\bibnamefont {Lo}}, \bibinfo {author} {\bibfnamefont {T.~G.~F.}\ \bibnamefont
			{Li}}, \ and\ \bibinfo {author} {\bibfnamefont {C.}~\bibnamefont {Van
				Den~Broeck}},\ }\href@noop {} {\  (\bibinfo {year} {2021})},\ \Eprint
	{http://arxiv.org/abs/2105.09485} {arXiv:2105.09485 [gr-qc]} \BibitemShut
	{NoStop}%
	\bibitem [{\citenamefont {Zhang}\ \emph {et~al.}(2022)\citenamefont {Zhang},
		\citenamefont {Gong}, \citenamefont {Liang},\ and\ \citenamefont
		{Zhang}}]{Zhang:2021fha}%
	\BibitemOpen
	\bibfield  {author} {\bibinfo {author} {\bibfnamefont {C.}~\bibnamefont
			{Zhang}}, \bibinfo {author} {\bibfnamefont {Y.}~\bibnamefont {Gong}},
		\bibinfo {author} {\bibfnamefont {D.}~\bibnamefont {Liang}}, \ and\ \bibinfo
		{author} {\bibfnamefont {C.}~\bibnamefont {Zhang}},\ }\href {\doibase
		10.1103/PhysRevD.105.104062} {\bibfield  {journal} {\bibinfo  {journal}
			{Phys. Rev. D}\ }\textbf {\bibinfo {volume} {105}},\ \bibinfo {pages}
		{104062} (\bibinfo {year} {2022})},\ \Eprint
	{http://arxiv.org/abs/2102.03972} {arXiv:2102.03972 [gr-qc]} \BibitemShut
	{NoStop}%
	\bibitem [{\citenamefont {Hu}\ \emph {et~al.}(2024)\citenamefont {Hu},
		\citenamefont {Liang},\ and\ \citenamefont {Shao}}]{Hu:2023soi}%
	\BibitemOpen
	\bibfield  {author} {\bibinfo {author} {\bibfnamefont {J.}~\bibnamefont
			{Hu}}, \bibinfo {author} {\bibfnamefont {D.}~\bibnamefont {Liang}}, \ and\
		\bibinfo {author} {\bibfnamefont {L.}~\bibnamefont {Shao}},\ }\href {\doibase
		10.1103/PhysRevD.109.084023} {\bibfield  {journal} {\bibinfo  {journal}
			{Phys. Rev. D}\ }\textbf {\bibinfo {volume} {109}},\ \bibinfo {pages}
		{084023} (\bibinfo {year} {2024})},\ \Eprint
	{http://arxiv.org/abs/2310.01249} {arXiv:2310.01249 [gr-qc]} \BibitemShut
	{NoStop}%
	\bibitem [{\citenamefont {Jiang}\ and\ \citenamefont
		{Zhang}(2026)}]{Jiang:2025abg}%
	\BibitemOpen
	\bibfield  {author} {\bibinfo {author} {\bibfnamefont {T.}~\bibnamefont
			{Jiang}}\ and\ \bibinfo {author} {\bibfnamefont {C.}~\bibnamefont {Zhang}},\
	}\href {\doibase 10.1140/epjc/s10052-026-15444-2} {\bibfield  {journal}
		{\bibinfo  {journal} {Eur. Phys. J. C}\ }\textbf {\bibinfo {volume} {86}},\
		\bibinfo {pages} {194} (\bibinfo {year} {2026})},\ \Eprint
	{http://arxiv.org/abs/2507.14870} {arXiv:2507.14870 [gr-qc]} \BibitemShut
	{NoStop}%
	\bibitem [{\citenamefont {Johnson-McDaniel}\ \emph {et~al.}(2022)\citenamefont
		{Johnson-McDaniel}, \citenamefont {Ghosh}, \citenamefont {Ghonge},
		\citenamefont {Saleem}, \citenamefont {Krishnendu},\ and\ \citenamefont
		{Clark}}]{Johnson-McDaniel:2021yge}%
	\BibitemOpen
	\bibfield  {author} {\bibinfo {author} {\bibfnamefont {N.~K.}\ \bibnamefont
			{Johnson-McDaniel}}, \bibinfo {author} {\bibfnamefont {A.}~\bibnamefont
			{Ghosh}}, \bibinfo {author} {\bibfnamefont {S.}~\bibnamefont {Ghonge}},
		\bibinfo {author} {\bibfnamefont {M.}~\bibnamefont {Saleem}}, \bibinfo
		{author} {\bibfnamefont {N.~V.}\ \bibnamefont {Krishnendu}}, \ and\ \bibinfo
		{author} {\bibfnamefont {J.~A.}\ \bibnamefont {Clark}},\ }\href {\doibase
		10.1103/PhysRevD.105.044020} {\bibfield  {journal} {\bibinfo  {journal}
			{Phys. Rev. D}\ }\textbf {\bibinfo {volume} {105}},\ \bibinfo {pages}
		{044020} (\bibinfo {year} {2022})},\ \Eprint
	{http://arxiv.org/abs/2109.06988} {arXiv:2109.06988 [gr-qc]} \BibitemShut
	{NoStop}%
	\bibitem [{\citenamefont {Nielsen}\ \emph {et~al.}(2019)\citenamefont
		{Nielsen}, \citenamefont {Nitz}, \citenamefont {Capano},\ and\ \citenamefont
		{Brown}}]{Nielsen:2018bhc}%
	\BibitemOpen
	\bibfield  {author} {\bibinfo {author} {\bibfnamefont {A.~B.}\ \bibnamefont
			{Nielsen}}, \bibinfo {author} {\bibfnamefont {A.~H.}\ \bibnamefont {Nitz}},
		\bibinfo {author} {\bibfnamefont {C.~D.}\ \bibnamefont {Capano}}, \ and\
		\bibinfo {author} {\bibfnamefont {D.~A.}\ \bibnamefont {Brown}},\ }\href
	{\doibase 10.1088/1475-7516/2019/02/019} {\bibfield  {journal} {\bibinfo
			{journal} {JCAP}\ }\textbf {\bibinfo {volume} {02}},\ \bibinfo {pages} {019}
		(\bibinfo {year} {2019})},\ \Eprint {http://arxiv.org/abs/1811.04071}
	{arXiv:1811.04071 [astro-ph.HE]} \BibitemShut {NoStop}%
	\bibitem [{\citenamefont {Marcoccia}\ \emph {et~al.}(2020)\citenamefont
		{Marcoccia}, \citenamefont {Fredriksson}, \citenamefont {Nielsen},\ and\
		\citenamefont {Nardini}}]{Marcoccia:2020rag}%
	\BibitemOpen
	\bibfield  {author} {\bibinfo {author} {\bibfnamefont {P.}~\bibnamefont
			{Marcoccia}}, \bibinfo {author} {\bibfnamefont {F.}~\bibnamefont
			{Fredriksson}}, \bibinfo {author} {\bibfnamefont {A.~B.}\ \bibnamefont
			{Nielsen}}, \ and\ \bibinfo {author} {\bibfnamefont {G.}~\bibnamefont
			{Nardini}},\ }\href {\doibase 10.1088/1475-7516/2020/11/043} {\bibfield
		{journal} {\bibinfo  {journal} {JCAP}\ }\textbf {\bibinfo {volume} {11}},\
		\bibinfo {pages} {043} (\bibinfo {year} {2020})},\ \Eprint
	{http://arxiv.org/abs/2008.12663} {arXiv:2008.12663 [gr-qc]} \BibitemShut
	{NoStop}%
	\bibitem [{\citenamefont {Liang}\ \emph
		{et~al.}(2026{\natexlab{a}})\citenamefont {Liang}, \citenamefont {Dai},\ and\
		\citenamefont {Yang}}]{Liang:2025zws}%
	\BibitemOpen
	\bibfield  {author} {\bibinfo {author} {\bibfnamefont {D.}~\bibnamefont
			{Liang}}, \bibinfo {author} {\bibfnamefont {N.}~\bibnamefont {Dai}}, \ and\
		\bibinfo {author} {\bibfnamefont {Y.}~\bibnamefont {Yang}},\ }\href {\doibase
		10.1088/1475-7516/2026/04/022} {\bibfield  {journal} {\bibinfo  {journal}
			{JCAP}\ }\textbf {\bibinfo {volume} {04}},\ \bibinfo {pages} {022} (\bibinfo
		{year} {2026}{\natexlab{a}})},\ \Eprint {http://arxiv.org/abs/2509.14924}
	{arXiv:2509.14924 [gr-qc]} \BibitemShut {NoStop}%
	\bibitem [{\citenamefont {Liang}\ \emph
		{et~al.}(2026{\natexlab{b}})\citenamefont {Liang}, \citenamefont {Wang},
		\citenamefont {Qin}, \citenamefont {Mai}, \citenamefont {Jiang},\ and\
		\citenamefont {Yang}}]{Liang:2026nij}%
	\BibitemOpen
	\bibfield  {author} {\bibinfo {author} {\bibfnamefont {D.}~\bibnamefont
			{Liang}}, \bibinfo {author} {\bibfnamefont {H.-T.}\ \bibnamefont {Wang}},
		\bibinfo {author} {\bibfnamefont {J.}~\bibnamefont {Qin}}, \bibinfo {author}
		{\bibfnamefont {Z.-F.}\ \bibnamefont {Mai}}, \bibinfo {author} {\bibfnamefont
			{T.}~\bibnamefont {Jiang}}, \ and\ \bibinfo {author} {\bibfnamefont
			{Y.}~\bibnamefont {Yang}},\ }\href@noop {} {\  (\bibinfo {year}
		{2026}{\natexlab{b}})},\ \Eprint {http://arxiv.org/abs/2608.06627}
	{arXiv:2608.06627 [gr-qc]} \BibitemShut {NoStop}%
	\bibitem [{\citenamefont {Abbott}\ \emph {et~al.}(2020)\citenamefont {Abbott}
		\emph {et~al.}}]{LIGOScientific:2019hgc}%
	\BibitemOpen
	\bibfield  {author} {\bibinfo {author} {\bibfnamefont {B.~P.}\ \bibnamefont
			{Abbott}} \emph {et~al.} (\bibinfo {collaboration} {LIGO Scientific,
			Virgo}),\ }\href {\doibase 10.1088/1361-6382/ab685e} {\bibfield  {journal}
		{\bibinfo  {journal} {Class. Quant. Grav.}\ }\textbf {\bibinfo {volume}
			{37}},\ \bibinfo {pages} {055002} (\bibinfo {year} {2020})},\ \Eprint
	{http://arxiv.org/abs/1908.11170} {arXiv:1908.11170 [gr-qc]} \BibitemShut
	{NoStop}%
	\bibitem [{\citenamefont {Davis}\ \emph {et~al.}(2021)\citenamefont {Davis}
		\emph {et~al.}}]{LIGO:2021ppb}%
	\BibitemOpen
	\bibfield  {author} {\bibinfo {author} {\bibfnamefont {D.}~\bibnamefont
			{Davis}} \emph {et~al.} (\bibinfo {collaboration} {LIGO}),\ }\href {\doibase
		10.1088/1361-6382/abfd85} {\bibfield  {journal} {\bibinfo  {journal} {Class.
				Quant. Grav.}\ }\textbf {\bibinfo {volume} {38}},\ \bibinfo {pages} {135014}
		(\bibinfo {year} {2021})},\ \Eprint {http://arxiv.org/abs/2101.11673}
	{arXiv:2101.11673 [astro-ph.IM]} \BibitemShut {NoStop}%
	\bibitem [{\citenamefont {Soni}\ \emph {et~al.}(2025)\citenamefont {Soni} \emph
		{et~al.}}]{LIGO:2024kkz}%
	\BibitemOpen
	\bibfield  {author} {\bibinfo {author} {\bibfnamefont {S.}~\bibnamefont
			{Soni}} \emph {et~al.} (\bibinfo {collaboration} {LIGO}),\ }\href {\doibase
		10.1088/1361-6382/adc4b6} {\bibfield  {journal} {\bibinfo  {journal} {Class.
				Quant. Grav.}\ }\textbf {\bibinfo {volume} {42}},\ \bibinfo {pages} {085016}
		(\bibinfo {year} {2025})},\ \Eprint {http://arxiv.org/abs/2409.02831}
	{arXiv:2409.02831 [astro-ph.IM]} \BibitemShut {NoStop}%
	\bibitem [{\citenamefont {Glanzer}\ \emph {et~al.}(2026)\citenamefont {Glanzer}
		\emph {et~al.}}]{LIGO:2026ika}%
	\BibitemOpen
	\bibfield  {author} {\bibinfo {author} {\bibfnamefont {J.}~\bibnamefont
			{Glanzer}} \emph {et~al.} (\bibinfo {collaboration} {LIGO}),\ }\href@noop {}
	{\  (\bibinfo {year} {2026})},\ \Eprint {http://arxiv.org/abs/2608.12193}
	{arXiv:2608.12193 [astro-ph.IM]} \BibitemShut {NoStop}%
	\bibitem [{\citenamefont {Abbott}\ \emph
		{et~al.}(2016{\natexlab{c}})\citenamefont {Abbott} \emph
		{et~al.}}]{LIGOScientific:2016gtq}%
	\BibitemOpen
	\bibfield  {author} {\bibinfo {author} {\bibfnamefont {B.~P.}\ \bibnamefont
			{Abbott}} \emph {et~al.} (\bibinfo {collaboration} {LIGO Scientific,
			Virgo}),\ }\href {\doibase 10.1088/0264-9381/33/13/134001} {\bibfield
		{journal} {\bibinfo  {journal} {Class. Quant. Grav.}\ }\textbf {\bibinfo
			{volume} {33}},\ \bibinfo {pages} {134001} (\bibinfo {year}
		{2016}{\natexlab{c}})},\ \Eprint {http://arxiv.org/abs/1602.03844}
	{arXiv:1602.03844 [gr-qc]} \BibitemShut {NoStop}%
	\bibitem [{\citenamefont {Zevin}\ \emph {et~al.}(2017)\citenamefont {Zevin}
		\emph {et~al.}}]{Zevin:2016qwy}%
	\BibitemOpen
	\bibfield  {author} {\bibinfo {author} {\bibfnamefont {M.}~\bibnamefont
			{Zevin}} \emph {et~al.},\ }\href {\doibase 10.1088/1361-6382/aa5cea}
	{\bibfield  {journal} {\bibinfo  {journal} {Class. Quant. Grav.}\ }\textbf
		{\bibinfo {volume} {34}},\ \bibinfo {pages} {064003} (\bibinfo {year}
		{2017})},\ \Eprint {http://arxiv.org/abs/1611.04596} {arXiv:1611.04596
		[gr-qc]} \BibitemShut {NoStop}%
	\bibitem [{\citenamefont {Glanzer}\ \emph {et~al.}(2023)\citenamefont {Glanzer}
		\emph {et~al.}}]{Glanzer:2022avx}%
	\BibitemOpen
	\bibfield  {author} {\bibinfo {author} {\bibfnamefont {J.}~\bibnamefont
			{Glanzer}} \emph {et~al.},\ }\href {\doibase 10.1088/1361-6382/acb633}
	{\bibfield  {journal} {\bibinfo  {journal} {Class. Quant. Grav.}\ }\textbf
		{\bibinfo {volume} {40}},\ \bibinfo {pages} {065004} (\bibinfo {year}
		{2023})},\ \Eprint {http://arxiv.org/abs/2208.12849} {arXiv:2208.12849
		[gr-qc]} \BibitemShut {NoStop}%
	\bibitem [{\citenamefont {Nuttall}(2018)}]{Nuttall:2018xhi}%
	\BibitemOpen
	\bibfield  {author} {\bibinfo {author} {\bibfnamefont {L.~K.}\ \bibnamefont
			{Nuttall}},\ }\href {\doibase 10.1098/rsta.2017.0286} {\bibfield  {journal}
		{\bibinfo  {journal} {Phil. Trans. Roy. Soc. Lond. A}\ }\textbf {\bibinfo
			{volume} {376}},\ \bibinfo {pages} {20170286} (\bibinfo {year} {2018})},\
	\Eprint {http://arxiv.org/abs/1804.07592} {arXiv:1804.07592 [astro-ph.IM]}
	\BibitemShut {NoStop}%
	\bibitem [{\citenamefont {Goetz}\ \emph {et~al.}(2026)\citenamefont {Goetz}
		\emph {et~al.}}]{O4LIGODetector:2026okh}%
	\BibitemOpen
	\bibfield  {author} {\bibinfo {author} {\bibfnamefont {E.}~\bibnamefont
			{Goetz}} \emph {et~al.} (\bibinfo {collaboration} {O4 LIGO Detector}),\
	}\href@noop {} {\  (\bibinfo {year} {2026})},\ \Eprint
	{http://arxiv.org/abs/2606.05959} {arXiv:2606.05959 [astro-ph.IM]}
	\BibitemShut {NoStop}%
	\bibitem [{\citenamefont {Gupte}\ \emph {et~al.}(2025)\citenamefont {Gupte}
		\emph {et~al.}}]{Gupte:2024jfe}%
	\BibitemOpen
	\bibfield  {author} {\bibinfo {author} {\bibfnamefont {N.}~\bibnamefont
			{Gupte}} \emph {et~al.},\ }\href {\doibase 10.1103/vpyp-nvfp} {\bibfield
		{journal} {\bibinfo  {journal} {Phys. Rev. D}\ }\textbf {\bibinfo {volume}
			{112}},\ \bibinfo {pages} {104045} (\bibinfo {year} {2025})},\ \Eprint
	{http://arxiv.org/abs/2404.14286} {arXiv:2404.14286 [gr-qc]} \BibitemShut
	{NoStop}%
	\bibitem [{\citenamefont {Jan}\ \emph {et~al.}(2026)\citenamefont {Jan},
		\citenamefont {Nicolella}, \citenamefont {Shoemaker},\ and\ \citenamefont
		{O'Shaughnessy}}]{Jan:2025zcm}%
	\BibitemOpen
	\bibfield  {author} {\bibinfo {author} {\bibfnamefont {A.}~\bibnamefont
			{Jan}}, \bibinfo {author} {\bibfnamefont {S.}~\bibnamefont {Nicolella}},
		\bibinfo {author} {\bibfnamefont {D.}~\bibnamefont {Shoemaker}}, \ and\
		\bibinfo {author} {\bibfnamefont {R.}~\bibnamefont {O'Shaughnessy}},\ }\href
	{\doibase 10.1103/ly9b-w75v} {\bibfield  {journal} {\bibinfo  {journal}
			{Phys. Rev. D}\ }\textbf {\bibinfo {volume} {113}},\ \bibinfo {pages}
		{084052} (\bibinfo {year} {2026})},\ \Eprint
	{http://arxiv.org/abs/2512.20060} {arXiv:2512.20060 [gr-qc]} \BibitemShut
	{NoStop}%
	\bibitem [{\citenamefont {Cole}\ \emph {et~al.}(2023)\citenamefont {Cole},
		\citenamefont {Coogan}, \citenamefont {Kavanagh},\ and\ \citenamefont
		{Bertone}}]{Cole:2022ucw}%
	\BibitemOpen
	\bibfield  {author} {\bibinfo {author} {\bibfnamefont {P.~S.}\ \bibnamefont
			{Cole}}, \bibinfo {author} {\bibfnamefont {A.}~\bibnamefont {Coogan}},
		\bibinfo {author} {\bibfnamefont {B.~J.}\ \bibnamefont {Kavanagh}}, \ and\
		\bibinfo {author} {\bibfnamefont {G.}~\bibnamefont {Bertone}},\ }\href
	{\doibase 10.1103/PhysRevD.107.083006} {\bibfield  {journal} {\bibinfo
			{journal} {Phys. Rev. D}\ }\textbf {\bibinfo {volume} {107}},\ \bibinfo
		{pages} {083006} (\bibinfo {year} {2023})},\ \Eprint
	{http://arxiv.org/abs/2207.07576} {arXiv:2207.07576 [astro-ph.CO]}
	\BibitemShut {NoStop}%
	\bibitem [{\citenamefont {Zwick}\ \emph {et~al.}(2024)\citenamefont {Zwick},
		\citenamefont {Tiede}, \citenamefont {Trani}, \citenamefont {Derdzinski},
		\citenamefont {Haiman}, \citenamefont {D'Orazio},\ and\ \citenamefont
		{Samsing}}]{Zwick:2024yzh}%
	\BibitemOpen
	\bibfield  {author} {\bibinfo {author} {\bibfnamefont {L.}~\bibnamefont
			{Zwick}}, \bibinfo {author} {\bibfnamefont {C.}~\bibnamefont {Tiede}},
		\bibinfo {author} {\bibfnamefont {A.~A.}\ \bibnamefont {Trani}}, \bibinfo
		{author} {\bibfnamefont {A.}~\bibnamefont {Derdzinski}}, \bibinfo {author}
		{\bibfnamefont {Z.}~\bibnamefont {Haiman}}, \bibinfo {author} {\bibfnamefont
			{D.~J.}\ \bibnamefont {D'Orazio}}, \ and\ \bibinfo {author} {\bibfnamefont
			{J.}~\bibnamefont {Samsing}},\ }\href {\doibase 10.1103/PhysRevD.110.103005}
	{\bibfield  {journal} {\bibinfo  {journal} {Phys. Rev. D}\ }\textbf {\bibinfo
			{volume} {110}},\ \bibinfo {pages} {103005} (\bibinfo {year} {2024})},\
	\Eprint {http://arxiv.org/abs/2405.05698} {arXiv:2405.05698 [gr-qc]}
	\BibitemShut {NoStop}%
	\bibitem [{\citenamefont {Zwick}\ \emph {et~al.}(2025)\citenamefont {Zwick},
		\citenamefont {Tak{\'a}tsy}, \citenamefont {Saini}, \citenamefont {Hendriks},
		\citenamefont {Samsing}, \citenamefont {Tiede}, \citenamefont {Rowan},\ and\
		\citenamefont {Trani}}]{Zwick:2025wkt}%
	\BibitemOpen
	\bibfield  {author} {\bibinfo {author} {\bibfnamefont {L.}~\bibnamefont
			{Zwick}}, \bibinfo {author} {\bibfnamefont {J.}~\bibnamefont {Tak{\'a}tsy}},
		\bibinfo {author} {\bibfnamefont {P.}~\bibnamefont {Saini}}, \bibinfo
		{author} {\bibfnamefont {K.}~\bibnamefont {Hendriks}}, \bibinfo {author}
		{\bibfnamefont {J.}~\bibnamefont {Samsing}}, \bibinfo {author} {\bibfnamefont
			{C.}~\bibnamefont {Tiede}}, \bibinfo {author} {\bibfnamefont
			{C.}~\bibnamefont {Rowan}}, \ and\ \bibinfo {author} {\bibfnamefont {A.~A.}\
			\bibnamefont {Trani}},\ }\href {\doibase 10.3847/1538-4357/adf6b8} {\bibfield
		{journal} {\bibinfo  {journal} {Astrophys. J.}\ }\textbf {\bibinfo {volume}
			{991}},\ \bibinfo {pages} {131} (\bibinfo {year} {2025})},\ \Eprint
	{http://arxiv.org/abs/2503.24084} {arXiv:2503.24084 [astro-ph.HE]}
	\BibitemShut {NoStop}%
	\bibitem [{\citenamefont {Samajdar}\ \emph {et~al.}(2021)\citenamefont
		{Samajdar}, \citenamefont {Janquart}, \citenamefont {Van Den~Broeck},\ and\
		\citenamefont {Dietrich}}]{Samajdar:2021egv}%
	\BibitemOpen
	\bibfield  {author} {\bibinfo {author} {\bibfnamefont {A.}~\bibnamefont
			{Samajdar}}, \bibinfo {author} {\bibfnamefont {J.}~\bibnamefont {Janquart}},
		\bibinfo {author} {\bibfnamefont {C.}~\bibnamefont {Van Den~Broeck}}, \ and\
		\bibinfo {author} {\bibfnamefont {T.}~\bibnamefont {Dietrich}},\ }\href
	{\doibase 10.1103/PhysRevD.104.044003} {\bibfield  {journal} {\bibinfo
			{journal} {Phys. Rev. D}\ }\textbf {\bibinfo {volume} {104}},\ \bibinfo
		{pages} {044003} (\bibinfo {year} {2021})},\ \Eprint
	{http://arxiv.org/abs/2102.07544} {arXiv:2102.07544 [gr-qc]} \BibitemShut
	{NoStop}%
	\bibitem [{\citenamefont {Relton}\ and\ \citenamefont
		{Raymond}(2021)}]{Relton:2021cax}%
	\BibitemOpen
	\bibfield  {author} {\bibinfo {author} {\bibfnamefont {P.}~\bibnamefont
			{Relton}}\ and\ \bibinfo {author} {\bibfnamefont {V.}~\bibnamefont
			{Raymond}},\ }\href {\doibase 10.1103/PhysRevD.104.084039} {\bibfield
		{journal} {\bibinfo  {journal} {Phys. Rev. D}\ }\textbf {\bibinfo {volume}
			{104}},\ \bibinfo {pages} {084039} (\bibinfo {year} {2021})},\ \Eprint
	{http://arxiv.org/abs/2103.16225} {arXiv:2103.16225 [gr-qc]} \BibitemShut
	{NoStop}%
	\bibitem [{\citenamefont {Powell}(2018)}]{Powell:2018csz}%
	\BibitemOpen
	\bibfield  {author} {\bibinfo {author} {\bibfnamefont {J.}~\bibnamefont
			{Powell}},\ }\href {\doibase 10.1088/1361-6382/aacf18} {\bibfield  {journal}
		{\bibinfo  {journal} {Class. Quant. Grav.}\ }\textbf {\bibinfo {volume}
			{35}},\ \bibinfo {pages} {155017} (\bibinfo {year} {2018})},\ \Eprint
	{http://arxiv.org/abs/1803.11346} {arXiv:1803.11346 [astro-ph.IM]}
	\BibitemShut {NoStop}%
	\bibitem [{\citenamefont {Hourihane}\ \emph {et~al.}(2022)\citenamefont
		{Hourihane}, \citenamefont {Chatziioannou}, \citenamefont {Wijngaarden},
		\citenamefont {Davis}, \citenamefont {Littenberg},\ and\ \citenamefont
		{Cornish}}]{Hourihane:2022doe}%
	\BibitemOpen
	\bibfield  {author} {\bibinfo {author} {\bibfnamefont {S.}~\bibnamefont
			{Hourihane}}, \bibinfo {author} {\bibfnamefont {K.}~\bibnamefont
			{Chatziioannou}}, \bibinfo {author} {\bibfnamefont {M.}~\bibnamefont
			{Wijngaarden}}, \bibinfo {author} {\bibfnamefont {D.}~\bibnamefont {Davis}},
		\bibinfo {author} {\bibfnamefont {T.}~\bibnamefont {Littenberg}}, \ and\
		\bibinfo {author} {\bibfnamefont {N.}~\bibnamefont {Cornish}},\ }\href
	{\doibase 10.1103/PhysRevD.106.042006} {\bibfield  {journal} {\bibinfo
			{journal} {Phys. Rev. D}\ }\textbf {\bibinfo {volume} {106}},\ \bibinfo
		{pages} {042006} (\bibinfo {year} {2022})},\ \Eprint
	{http://arxiv.org/abs/2205.13580} {arXiv:2205.13580 [gr-qc]} \BibitemShut
	{NoStop}%
	\bibitem [{\citenamefont {Narola}\ \emph {et~al.}(2025)\citenamefont {Narola}
		\emph {et~al.}}]{Narola:2024qdh}%
	\BibitemOpen
	\bibfield  {author} {\bibinfo {author} {\bibfnamefont {H.}~\bibnamefont
			{Narola}} \emph {et~al.},\ }\href {\doibase 10.1103/l6tp-ykxp} {\bibfield
		{journal} {\bibinfo  {journal} {Phys. Rev. D}\ }\textbf {\bibinfo {volume}
			{112}},\ \bibinfo {pages} {024079} (\bibinfo {year} {2025})},\ \Eprint
	{http://arxiv.org/abs/2411.15506} {arXiv:2411.15506 [gr-qc]} \BibitemShut
	{NoStop}%
	\bibitem [{\citenamefont {Cornish}\ and\ \citenamefont
		{Littenberg}(2015)}]{Cornish:2014kda}%
	\BibitemOpen
	\bibfield  {author} {\bibinfo {author} {\bibfnamefont {N.~J.}\ \bibnamefont
			{Cornish}}\ and\ \bibinfo {author} {\bibfnamefont {T.~B.}\ \bibnamefont
			{Littenberg}},\ }\href {\doibase 10.1088/0264-9381/32/13/135012} {\bibfield
		{journal} {\bibinfo  {journal} {Class. Quant. Grav.}\ }\textbf {\bibinfo
			{volume} {32}},\ \bibinfo {pages} {135012} (\bibinfo {year} {2015})},\
	\Eprint {http://arxiv.org/abs/1410.3835} {arXiv:1410.3835 [gr-qc]}
	\BibitemShut {NoStop}%
	\bibitem [{\citenamefont {Cornish}\ \emph {et~al.}(2021)\citenamefont
		{Cornish}, \citenamefont {Littenberg}, \citenamefont {B{\'e}csy},
		\citenamefont {Chatziioannou}, \citenamefont {Clark}, \citenamefont
		{Ghonge},\ and\ \citenamefont {Millhouse}}]{Cornish:2020dwh}%
	\BibitemOpen
	\bibfield  {author} {\bibinfo {author} {\bibfnamefont {N.~J.}\ \bibnamefont
			{Cornish}}, \bibinfo {author} {\bibfnamefont {T.~B.}\ \bibnamefont
			{Littenberg}}, \bibinfo {author} {\bibfnamefont {B.}~\bibnamefont
			{B{\'e}csy}}, \bibinfo {author} {\bibfnamefont {K.}~\bibnamefont
			{Chatziioannou}}, \bibinfo {author} {\bibfnamefont {J.~A.}\ \bibnamefont
			{Clark}}, \bibinfo {author} {\bibfnamefont {S.}~\bibnamefont {Ghonge}}, \
		and\ \bibinfo {author} {\bibfnamefont {M.}~\bibnamefont {Millhouse}},\ }\href
	{\doibase 10.1103/PhysRevD.103.044006} {\bibfield  {journal} {\bibinfo
			{journal} {Phys. Rev. D}\ }\textbf {\bibinfo {volume} {103}},\ \bibinfo
		{pages} {044006} (\bibinfo {year} {2021})},\ \Eprint
	{http://arxiv.org/abs/2011.09494} {arXiv:2011.09494 [gr-qc]} \BibitemShut
	{NoStop}%
	\bibitem [{\citenamefont {Pratten}\ \emph {et~al.}(2021)\citenamefont {Pratten}
		\emph {et~al.}}]{Pratten:2020ceb}%
	\BibitemOpen
	\bibfield  {author} {\bibinfo {author} {\bibfnamefont {G.}~\bibnamefont
			{Pratten}} \emph {et~al.},\ }\href {\doibase 10.1103/PhysRevD.103.104056}
	{\bibfield  {journal} {\bibinfo  {journal} {Phys. Rev. D}\ }\textbf {\bibinfo
			{volume} {103}},\ \bibinfo {pages} {104056} (\bibinfo {year} {2021})},\
	\Eprint {http://arxiv.org/abs/2004.06503} {arXiv:2004.06503 [gr-qc]}
	\BibitemShut {NoStop}%
	\bibitem [{\citenamefont {Ramos-Buades}\ \emph {et~al.}(2023)\citenamefont
		{Ramos-Buades}, \citenamefont {Buonanno}, \citenamefont {Estell{\'e}s},
		\citenamefont {Khalil}, \citenamefont {Mihaylov}, \citenamefont {Ossokine},
		\citenamefont {Pompili},\ and\ \citenamefont
		{Shiferaw}}]{Ramos-Buades:2023ehm}%
	\BibitemOpen
	\bibfield  {author} {\bibinfo {author} {\bibfnamefont {A.}~\bibnamefont
			{Ramos-Buades}}, \bibinfo {author} {\bibfnamefont {A.}~\bibnamefont
			{Buonanno}}, \bibinfo {author} {\bibfnamefont {H.}~\bibnamefont
			{Estell{\'e}s}}, \bibinfo {author} {\bibfnamefont {M.}~\bibnamefont
			{Khalil}}, \bibinfo {author} {\bibfnamefont {D.~P.}\ \bibnamefont
			{Mihaylov}}, \bibinfo {author} {\bibfnamefont {S.}~\bibnamefont {Ossokine}},
		\bibinfo {author} {\bibfnamefont {L.}~\bibnamefont {Pompili}}, \ and\
		\bibinfo {author} {\bibfnamefont {M.}~\bibnamefont {Shiferaw}},\ }\href
	{\doibase 10.1103/PhysRevD.108.124037} {\bibfield  {journal} {\bibinfo
			{journal} {Phys. Rev. D}\ }\textbf {\bibinfo {volume} {108}},\ \bibinfo
		{pages} {124037} (\bibinfo {year} {2023})},\ \Eprint
	{http://arxiv.org/abs/2303.18046} {arXiv:2303.18046 [gr-qc]} \BibitemShut
	{NoStop}%
	\bibitem [{\citenamefont {Trovato}(2020)}]{Trovato:2019liz}%
	\BibitemOpen
	\bibfield  {author} {\bibinfo {author} {\bibfnamefont {A.}~\bibnamefont
			{Trovato}} (\bibinfo {collaboration} {Ligo Scientific, Virgo}),\ }\href
	{\doibase 10.22323/1.357.0082} {\bibfield  {journal} {\bibinfo  {journal}
			{PoS}\ }\textbf {\bibinfo {volume} {Asterics2019}},\ \bibinfo {pages} {082}
		(\bibinfo {year} {2020})}\BibitemShut {NoStop}%
	\bibitem [{\citenamefont {Macleod}\ \emph {et~al.}(2021)\citenamefont
		{Macleod}, \citenamefont {Areeda}, \citenamefont {Coughlin}, \citenamefont
		{Massinger},\ and\ \citenamefont {Urban}}]{MACLEOD2021100657}%
	\BibitemOpen
	\bibfield  {author} {\bibinfo {author} {\bibfnamefont {D.~M.}\ \bibnamefont
			{Macleod}}, \bibinfo {author} {\bibfnamefont {J.~S.}\ \bibnamefont {Areeda}},
		\bibinfo {author} {\bibfnamefont {S.~B.}\ \bibnamefont {Coughlin}}, \bibinfo
		{author} {\bibfnamefont {T.~J.}\ \bibnamefont {Massinger}}, \ and\ \bibinfo
		{author} {\bibfnamefont {A.~L.}\ \bibnamefont {Urban}},\ }\href {\doibase
		https://doi.org/10.1016/j.softx.2021.100657} {\bibfield  {journal} {\bibinfo
			{journal} {SoftwareX}\ }\textbf {\bibinfo {volume} {13}},\ \bibinfo {pages}
		{100657} (\bibinfo {year} {2021})}\BibitemShut {NoStop}%
	\bibitem [{\citenamefont {Chatterji}\ \emph {et~al.}(2004)\citenamefont
		{Chatterji}, \citenamefont {Blackburn}, \citenamefont {Martin},\ and\
		\citenamefont {Katsavounidis}}]{Chatterji:2004qg}%
	\BibitemOpen
	\bibfield  {author} {\bibinfo {author} {\bibfnamefont {S.}~\bibnamefont
			{Chatterji}}, \bibinfo {author} {\bibfnamefont {L.}~\bibnamefont
			{Blackburn}}, \bibinfo {author} {\bibfnamefont {G.}~\bibnamefont {Martin}}, \
		and\ \bibinfo {author} {\bibfnamefont {E.}~\bibnamefont {Katsavounidis}},\
	}\href {\doibase 10.1088/0264-9381/21/20/024} {\bibfield  {journal} {\bibinfo
			{journal} {Class. Quant. Grav.}\ }\textbf {\bibinfo {volume} {21}},\
		\bibinfo {pages} {S1809} (\bibinfo {year} {2004})},\ \Eprint
	{http://arxiv.org/abs/gr-qc/0412119} {arXiv:gr-qc/0412119} \BibitemShut
	{NoStop}%
	\bibitem [{\citenamefont {Vazsonyi}\ and\ \citenamefont
		{Davis}(2023)}]{Vazsonyi:2022jul}%
	\BibitemOpen
	\bibfield  {author} {\bibinfo {author} {\bibfnamefont {L.}~\bibnamefont
			{Vazsonyi}}\ and\ \bibinfo {author} {\bibfnamefont {D.}~\bibnamefont
			{Davis}},\ }\href {\doibase 10.1088/1361-6382/acafd2} {\bibfield  {journal}
		{\bibinfo  {journal} {Class. Quant. Grav.}\ }\textbf {\bibinfo {volume}
			{40}},\ \bibinfo {pages} {035008} (\bibinfo {year} {2023})},\ \Eprint
	{http://arxiv.org/abs/2208.12338} {arXiv:2208.12338 [astro-ph.IM]}
	\BibitemShut {NoStop}%
	\bibitem [{\citenamefont {Chatterji}(2005)}]{Chatterji}%
	\BibitemOpen
	\bibfield  {author} {\bibinfo {author} {\bibfnamefont {S.~K.}\ \bibnamefont
			{Chatterji}},\ }\emph {\bibinfo {title} {The Search for Gravitational Wave
			Bursts in Data from the Second LIGO Science Run}},\ \href
	{https://hdl.handle.net/1721.1/34388} {Ph.D. thesis},\ \bibinfo  {school}
	{Massachusetts Institute of Technology} (\bibinfo {year} {2005})\BibitemShut
	{NoStop}%
	\bibitem [{\citenamefont {Blackburn}\ \emph {et~al.}(2008)\citenamefont
		{Blackburn} \emph {et~al.}}]{Blackburn:2008ah}%
	\BibitemOpen
	\bibfield  {author} {\bibinfo {author} {\bibfnamefont {L.}~\bibnamefont
			{Blackburn}} \emph {et~al.},\ }\href {\doibase
		10.1088/0264-9381/25/18/184004} {\bibfield  {journal} {\bibinfo  {journal}
			{Class. Quant. Grav.}\ }\textbf {\bibinfo {volume} {25}},\ \bibinfo {pages}
		{184004} (\bibinfo {year} {2008})},\ \Eprint {http://arxiv.org/abs/0804.0800}
	{arXiv:0804.0800 [gr-qc]} \BibitemShut {NoStop}%
	\bibitem [{\citenamefont {Kolmogorov}(1933)}]{Kolmogorov}%
	\BibitemOpen
	\bibfield  {author} {\bibinfo {author} {\bibfnamefont {A.~N.}\ \bibnamefont
			{Kolmogorov}},\ }\href@noop {} {\bibfield  {journal} {\bibinfo  {journal}
			{Giornale dell'Istituto Italiano degli Attuari}\ }\textbf {\bibinfo {volume}
			{4}},\ \bibinfo {pages} {83} (\bibinfo {year} {1933})}\BibitemShut {NoStop}%
	\bibitem [{\citenamefont {Smirnov}(1948)}]{N.-Smirnov}%
	\BibitemOpen
	\bibfield  {author} {\bibinfo {author} {\bibfnamefont {N.}~\bibnamefont
			{Smirnov}},\ }\href {http://www.jstor.org/stable/2236278} {\bibfield
		{journal} {\bibinfo  {journal} {The Annals of Mathematical Statistics}\
		}\textbf {\bibinfo {volume} {19}},\ \bibinfo {pages} {279} (\bibinfo {year}
		{1948})}\BibitemShut {NoStop}%
	\bibitem [{\citenamefont {Anderson}\ and\ \citenamefont
		{Darling}(1954)}]{T.-W.-Anderson}%
	\BibitemOpen
	\bibfield  {author} {\bibinfo {author} {\bibfnamefont {T.~W.}\ \bibnamefont
			{Anderson}}\ and\ \bibinfo {author} {\bibfnamefont {D.~A.}\ \bibnamefont
			{Darling}},\ }\href {http://www.jstor.org/stable/2281537} {\bibfield
		{journal} {\bibinfo  {journal} {Journal of the American Statistical
				Association}\ }\textbf {\bibinfo {volume} {49}},\ \bibinfo {pages} {765}
		(\bibinfo {year} {1954})}\BibitemShut {NoStop}%
	\bibitem [{\citenamefont {Pearson}(1900)}]{pearson1900x}%
	\BibitemOpen
	\bibfield  {author} {\bibinfo {author} {\bibfnamefont {K.}~\bibnamefont
			{Pearson}},\ }\href@noop {} {\bibfield  {journal} {\bibinfo  {journal} {The
				London, Edinburgh, and Dublin Philosophical Magazine and Journal of Science}\
		}\textbf {\bibinfo {volume} {50}},\ \bibinfo {pages} {157} (\bibinfo {year}
		{1900})}\BibitemShut {NoStop}%
	\bibitem [{\citenamefont {Abac}\ \emph
		{et~al.}(2025{\natexlab{b}})\citenamefont {Abac} \emph
		{et~al.}}]{LIGOScientific:2025rid}%
	\BibitemOpen
	\bibfield  {author} {\bibinfo {author} {\bibfnamefont {A.~G.}\ \bibnamefont
			{Abac}} \emph {et~al.} (\bibinfo {collaboration} {LIGO Scientific, Virgo,
			KAGRA}),\ }\href {\doibase 10.1103/kw5g-d732} {\bibfield  {journal} {\bibinfo
			{journal} {Phys. Rev. Lett.}\ }\textbf {\bibinfo {volume} {135}},\ \bibinfo
		{pages} {111403} (\bibinfo {year} {2025}{\natexlab{b}})},\ \Eprint
	{http://arxiv.org/abs/2509.08054} {arXiv:2509.08054 [gr-qc]} \BibitemShut
	{NoStop}%
	\bibitem [{\citenamefont {Fisher}(1932)}]{fisher1932statistical}%
	\BibitemOpen
	\bibfield  {author} {\bibinfo {author} {\bibfnamefont {R.}~\bibnamefont
			{Fisher}},\ }\href {https://books.google.com/books?id=cEwNAQAAIAAJ} {\emph
		{\bibinfo {title} {Statistical Methods for Research Workers}}},\ Biological
	monographs and manuals\ (\bibinfo  {publisher} {Oliver and Boyd},\ \bibinfo
	{year} {1932})\BibitemShut {NoStop}%
	\bibitem [{\citenamefont {Varma}\ \emph {et~al.}(2019)\citenamefont {Varma},
		\citenamefont {Field}, \citenamefont {Scheel}, \citenamefont {Blackman},
		\citenamefont {Gerosa}, \citenamefont {Stein}, \citenamefont {Kidder},\ and\
		\citenamefont {Pfeiffer}}]{Varma:2019csw}%
	\BibitemOpen
	\bibfield  {author} {\bibinfo {author} {\bibfnamefont {V.}~\bibnamefont
			{Varma}}, \bibinfo {author} {\bibfnamefont {S.~E.}\ \bibnamefont {Field}},
		\bibinfo {author} {\bibfnamefont {M.~A.}\ \bibnamefont {Scheel}}, \bibinfo
		{author} {\bibfnamefont {J.}~\bibnamefont {Blackman}}, \bibinfo {author}
		{\bibfnamefont {D.}~\bibnamefont {Gerosa}}, \bibinfo {author} {\bibfnamefont
			{L.~C.}\ \bibnamefont {Stein}}, \bibinfo {author} {\bibfnamefont {L.~E.}\
			\bibnamefont {Kidder}}, \ and\ \bibinfo {author} {\bibfnamefont {H.~P.}\
			\bibnamefont {Pfeiffer}},\ }\href {\doibase 10.1103/PhysRevResearch.1.033015}
	{\bibfield  {journal} {\bibinfo  {journal} {Phys. Rev. Research.}\ }\textbf
		{\bibinfo {volume} {1}},\ \bibinfo {pages} {033015} (\bibinfo {year}
		{2019})},\ \Eprint {http://arxiv.org/abs/1905.09300} {arXiv:1905.09300
		[gr-qc]} \BibitemShut {NoStop}%
	\bibitem [{\citenamefont {Rousseeuw}\ and\ \citenamefont
		{Croux}(1993)}]{Peter_J}%
	\BibitemOpen
	\bibfield  {author} {\bibinfo {author} {\bibfnamefont {P.~J.}\ \bibnamefont
			{Rousseeuw}}\ and\ \bibinfo {author} {\bibfnamefont {C.}~\bibnamefont
			{Croux}},\ }\href {http://www.jstor.org/stable/2291267} {\bibfield  {journal}
		{\bibinfo  {journal} {Journal of the American Statistical Association}\
		}\textbf {\bibinfo {volume} {88}},\ \bibinfo {pages} {1273} (\bibinfo {year}
		{1993})}\BibitemShut {NoStop}%
	\bibitem [{\citenamefont {Punturo}\ \emph {et~al.}(2010)\citenamefont {Punturo}
		\emph {et~al.}}]{Punturo:2010zz}%
	\BibitemOpen
	\bibfield  {author} {\bibinfo {author} {\bibfnamefont {M.}~\bibnamefont
			{Punturo}} \emph {et~al.},\ }\href {\doibase 10.1088/0264-9381/27/19/194002}
	{\bibfield  {journal} {\bibinfo  {journal} {Class. Quant. Grav.}\ }\textbf
		{\bibinfo {volume} {27}},\ \bibinfo {pages} {194002} (\bibinfo {year}
		{2010})}\BibitemShut {NoStop}%
	\bibitem [{\citenamefont {Maggiore}\ \emph {et~al.}(2020)\citenamefont
		{Maggiore} \emph {et~al.}}]{ET:2019dnz}%
	\BibitemOpen
	\bibfield  {author} {\bibinfo {author} {\bibfnamefont {M.}~\bibnamefont
			{Maggiore}} \emph {et~al.} (\bibinfo {collaboration} {ET}),\ }\href {\doibase
		10.1088/1475-7516/2020/03/050} {\bibfield  {journal} {\bibinfo  {journal}
			{JCAP}\ }\textbf {\bibinfo {volume} {03}},\ \bibinfo {pages} {050} (\bibinfo
		{year} {2020})},\ \Eprint {http://arxiv.org/abs/1912.02622} {arXiv:1912.02622
		[astro-ph.CO]} \BibitemShut {NoStop}%
	\bibitem [{\citenamefont {Reitze}\ \emph {et~al.}(2019)\citenamefont {Reitze}
		\emph {et~al.}}]{Reitze:2019iox}%
	\BibitemOpen
	\bibfield  {author} {\bibinfo {author} {\bibfnamefont {D.}~\bibnamefont
			{Reitze}} \emph {et~al.},\ }\href@noop {} {\bibfield  {journal} {\bibinfo
			{journal} {Bull. Am. Astron. Soc.}\ }\textbf {\bibinfo {volume} {51}},\
		\bibinfo {pages} {035} (\bibinfo {year} {2019})},\ \Eprint
	{http://arxiv.org/abs/1907.04833} {arXiv:1907.04833 [astro-ph.IM]}
	\BibitemShut {NoStop}%
	\bibitem [{\citenamefont {Hild}\ \emph {et~al.}(2011)\citenamefont {Hild} \emph
		{et~al.}}]{Hild:2010id}%
	\BibitemOpen
	\bibfield  {author} {\bibinfo {author} {\bibfnamefont {S.}~\bibnamefont
			{Hild}} \emph {et~al.},\ }\href {\doibase 10.1088/0264-9381/28/9/094013}
	{\bibfield  {journal} {\bibinfo  {journal} {Class. Quant. Grav.}\ }\textbf
		{\bibinfo {volume} {28}},\ \bibinfo {pages} {094013} (\bibinfo {year}
		{2011})},\ \Eprint {http://arxiv.org/abs/1012.0908} {arXiv:1012.0908 [gr-qc]}
	\BibitemShut {NoStop}%
\end{thebibliography}


\end{document}